\documentclass[aps,prx,twocolumn,superscriptaddress,showpacs,floatfix,longbibliography]{revtex4-2}
\usepackage{graphicx} % Required for inserting images
\usepackage{amsmath,amssymb,amsfonts,float,graphics,epsfig,epstopdf,color,verbatim,tabularx,bm,multirow,appendix}
\usepackage{tikz}
\usetikzlibrary{arrows.meta,positioning}
\usepackage{amsmath,amssymb,graphicx}
\usepackage{wasysym}
\usepackage[utf8]{inputenc}
\usepackage[T1]{fontenc}
\usepackage{xcolor}
\usepackage{dsfont}
\usepackage{textcomp}
\usepackage{bm}

\usepackage{graphicx}% 
\usepackage{xcolor}
\usepackage{dcolumn}
\usepackage{bm}
\usepackage{xspace}
\usepackage{time}
\usepackage{booktabs}
\usepackage{multirow}
\usepackage{makecell}
\usepackage[normalem]{ulem} % Use [normalem] to prevent changing the behavior of \emph{} (italics)

\pdfoutput=1
\usepackage{color}
\definecolor{LinkColor}{rgb}{0.256,0.439,0.588}
\usepackage{hyperref}
\hypersetup{
colorlinks=true,
citecolor=LinkColor,
linkcolor=LinkColor,
urlcolor=LinkColor
}
\usepackage{cleveref}
\Crefname{equation}{Eq.}{Eqs.}
\Crefname{figure}{Fig.}{Figs.}
\newcommand{\di}{\mathrm{d}}
\newcommand{\odiff}[2]{\frac{\di #1}{\di #2}}
\newcommand{\pdiff}[2]{\frac{\partial #1}{\partial #2}}

\newcommand{\ket}[1]{\left \lvert#1\right \rangle}
\newcommand{\overlap}[2]{\langle#1\vert#2\rangle}

\graphicspath{{./figures/}{./trees/}}

\begin{document}
\title{Hybrid Monte Carlo for Fractional Quantum Hall States}
\author{Ting-Tung Wang}
\affiliation{Department of Physics and HK Institute of Quantum Science \& Technology, The University of Hong Kong, Pokfulam Road,  Hong Kong SAR, China}
\affiliation{State Key Laboratory of Optical Quantum Materials, The University of Hong Kong, Pokfulam Road,  Hong Kong SAR, China}

\author{Ha Quang Trung}
\affiliation{Division of Physics and Applied Physics, Nanyang Technological University, Singapore 637371, Singapore}

\author{Qianhui Xu}
\affiliation{Division of Physics and Applied Physics, Nanyang Technological University, Singapore 637371, Singapore}

\author{Min Long}
\affiliation{Department of Physics and HK Institute of Quantum Science \& Technology, The University of Hong Kong, Pokfulam Road,  Hong Kong SAR, China}
\affiliation{State Key Laboratory of Optical Quantum Materials, The University of Hong Kong, Pokfulam Road,  Hong Kong SAR, China}

\author{Bo Yang}
\email{yang.bo@ntu.edu.sg}
\affiliation{Division of Physics and Applied Physics, Nanyang Technological University, Singapore 637371, Singapore}

\author{Zi Yang Meng}
\email{zymeng@hku.hk}
\affiliation{Department of Physics and HK Institute of Quantum Science \& Technology, The University of Hong Kong, Pokfulam Road,  Hong Kong SAR, China}
\affiliation{State Key Laboratory of Optical Quantum Materials, The University of Hong Kong, Pokfulam Road,  Hong Kong SAR, China}

\date{\today}

\begin{abstract}
We develop a hybrid Monte Carlo method to efficiently compute the physical observables from the samplings of the Laughlin and the Moore-Read wave functions of fractional quantum Hall (FQH) systems. With the advancements in methodology, including global updates and double stereographic projection on spherical geometry, our hybrid Monte Carlo simulation is significantly faster than the widely used Metropolis Monte Carlo scheme. As a result, we can readily simulate systems with electron numbers $N > 1000$ on both disk and sphere geometries. We apply this method to investigating the topological shift obtained from the edge dipole moment, computed from the density of the wave function on the disk. We also numerically computed the non-Abelian braiding matrices for different braiding schemes of the Moore-Read quasiholes on the sphere. Results with much better quality compared with previous works have been achieved. With the thermodynamic limit results obtained at ease, we also discuss the future usage of our method to clarify the questions on the instability of fractional quantum Hall states in an ideal Chern band setting or under quantum decoherence.
\end{abstract}

\maketitle

\section{Introduction}
The fractional quantum Hall (FQH) liquids that emerge in a two-dimensional electron gas subjected to a strong perpendicular magnetic field at low temperatures host intrinsic topological orders and quasiparticle excitations that are anyons with fractional charge and fractional statistics~\cite{tsui1982two,laughlin1983,halperin1986calculations,arovas1984fractional,prange1987quantum,feldman2021fractional}. The universal properties of FQH ground states and their excitations are encoded in topological invariants such as the Hall conductivity, the topological shift, and the braiding statistics of quasihole excitations~\cite{laughlin1983,halperin1986calculations,arovas1984fractional,feldman2021fractional}. Establishing and quantitatively characterizing such topological data in the thermodynamic limit is therefore a central physics problem, both for a fundamental understanding of topological phases and for the engineering applications, such as in fault-tolerant topological quantum computation~\cite{preskill2004topological}

Addressing these questions quantitatively requires numerical access to large systems. 
To date, numerical studies of anyonic properties have mostly used exact diagonalization (ED), Metropolis Monte Carlo sampling with local updates of electron coordinates, and matrix product state (MPS) representations of FQH wave functions on disk, cylinder, or sphere geometries~\cite{morfMonte1986,melikQuantum1997,tserkovnyak2003monte,ciftjaMonte2003,Zaletel2012Exact,wu2014braiding,umucalilar2018time,macaluso2019fusion,macalusoCharge2020,comparin2022measurable,nardinSpin2023,jiUniversal2024,trung2025long,fagerlundSpin2025,xuDynamics2025,boseMonteCarlo2025,chenProbing2025}. 
However, each of these approaches has its own limitation: ED suffers from the exponential growth of the Hilbert space~\cite{Prodan_Mapping_2009,li2022anyonic}, conventional Metropolis updates are strictly local and thus can exhibit long autocorrelation times, and MPS calculations on cylinders may be influenced by gapless edge modes while requiring rapidly increasing dimension as the number of quasiholes grows. 

These limitations become especially acute for non-Abelian phases, among which a prime example is the filling $\nu=5/2$ plateau, widely believed to realize the Moore-Read (MR) phase supporting the emergence of non-Abelian anyons~\cite{moore1991nonabelions,nayak19962n,bonderson2006detecting,bonderson2008interferometry, Bonderson2011}. 
Demonstrating the existence of these quasiparticles and quantifying their robustness are central to both fundamental theories and topological quantum computation proposals, in which braiding implements unitary gates that are intrinsically protected against local perturbations and decoherence~\cite{kitaev2003fault,freedman2006towards,nayakNonAbelian2008,wangFractional2025}. 
Although numerical studies of the ideal MR model reported quasihole braiding consistent with Ising non-Abelian statistics and spectra compatible with an Ising$\times$U(1) CFT structure~\cite{moore1989classical,moore1991nonabelions,gurariePlasma1997,nayakNonAbelian2008}, controlled estimates of more delicate properties -- including excitation gaps, stability under realistic perturbations and decoherence, and braiding -- remain challenging because of the strong finite-size effect, large quantum oscillation, and the strict requirement of adiabatic transformation ~\cite{tserkovnyak2003monte,comparin2022measurable,trung2023spin,macaluso2019fusion}. 

From a computational standpoint, the difficulty is amplified by the structure of the MR wave functions: Monte Carlo sampling typically requires repeated evaluation of Pfaffians and/or determinants, whose direct updates scale as $O(N^3)$, where $N$ is the number of electrons in the many-body wave function, making large-$N$ simulations expensive.
When this cost is combined with local random-walk moves in standard Metropolis schemes, the resulting long autocorrelation times have largely confined non-Abelian Monte Carlo studies to $N\lesssim 10^2$, far below the $N\sim 10^3$ scales routinely reached in other strongly correlated settings (e.g., Hubbard-model simulations)~\cite{xuRevealing2019}. 
Consequently, a key open technical need is a sampling strategy that simultaneously (i) performs global, geometry-respecting updates, and (ii) remains computationally scalable at large $N$.

In this work, we address these challenges by developing a hybrid Monte Carlo (HMC) framework for FQH trial wave functions both on the disk and spherical geometry. 
In particular, by incorporating a double stereographic projection in spherical geometry, our method enables genuinely global updates of electron coordinates generated by Hamiltonian-guided dynamics, substantially reducing autocorrelation and improving sampling efficiency compared with local Metropolis updates. 
This advance allows us to easily reach $N>1000$ with moderate computational resources and to extract anyon-related topological data in the thermodynamic regime. 
We determine the topological shift for Laughlin and Moore-Read states and, more importantly, compute the non-Abelian braiding matrix of braiding schemes within the two-fold degenerate four-quasihole MR manifold. Our results present better data quality compared to previous works; each measurement finishes within 7 days for the Laughlin state (up to $N = 1200$) and MR state (up to $N = 400$) and on a single compute node (with 32 Intel Xeon Gold 6226R CPU cores).

With the thermodynamic limit results obtained at ease, our method can be used to clarify the recent questions on the instability of fractional quantum Hall states in an ideal Chern band setting with a nonuniform magnetic field~\cite{moitraInstability2025} or under density decoherence~\cite{wangFractional2025}. In both cases, there are proposals that the FQH liquids (both Laughlin and Moore-Read) are not stable towards gapless states through the Berezinskii-Kosterlitz-Thouless (BKT) transition. We foresee that our hybrid Monte Carlo scheme will have the assessment of the necessary system sizes~\cite{jiangMonte2022,jiangSolving2022} to clarify the situation.

%that can unambiguously capture the associated essential singularity and power-law correlations with changing exponents~\cite{jiangMonte2022,jiangSolving2022} and eventually clarify the situation.

The rest of the paper is organized as follows. Sec.~\ref{sec:II} outlines our hybrid Monte Carlo method for both disk and sphere geometries using the Laughlin wave function as an example. We also explain why our method has superior performance compared with the traditional Metropolis Monte Carlo. Sec.~\ref{sec:III} presents our main results, with Sec.~\ref{sec:IIIA} discussing the topological shift in the Laughlin wave function on the disk and Secs.~\ref{sec:IIIB} and \ref{sec:IIIC} presenting the data for the Berry phase and braiding matrix for the MR states under our double stereographic projection on the sphere. We then summarize and present the outline for future works in Sec.~\ref{sec:IV}. Detailed derivation of the Berry matrix, fusion and braiding in anyon models, and the double stereographic projection for our HMC simulation on the sphere, are given in the Supplementary Material (SM)~\cite{suppl}.

\section{Hybrid Monte Carlo Method on FQH wave functions}
\label{sec:II}
The Laughlin wave function provides a paradigmatic description of the FQH states at filling factor $\nu = 1/m$, where $m$ is a positive integer ~\cite{laughlin1983,haldaneFractional1983}. For $N$ electrons confined to the lowest Landau level (LLL) in a two-dimensional plane under a perpendicular magnetic field, the Laughlin state is expressed in complex coordinates $z_j = x_j + i y_j$ (with the magnetic length set to unity, $\ell_B = 1$) as
\begin{equation}
\Psi_m(\{z_j\}) = \prod_{i<j}^N (z_i - z_j)^m \exp\left( -\sum_k |z_k|^2 / 4 \right),
\end{equation}
up to a normalization constant. This many-body wave function is an exact eigenstate of a short-range repulsive interaction and exhibits non-trivial topological order, characterized by fractional statistics of quasiparticles ~\cite{laughlin1983,haldaneFractional1983,halperin1986calculations,feldman2021fractional}.

Direct analytical evaluation of physical observables in the Laughlin state is challenging due to the $2N$-dimensional configuration space. However, it is crucial to study the behavior of many-body wave functions with a large number of particles, especially for the edge dipole moment and braiding of quasiholes, where the bulk region of the disk and between quasi-particles need to be large to overcome the finite-size effect~\cite{park2014guiding,trung2023spin,jiUniversal2024}. Monte Carlo (MC) method~\cite{metropolisEquation1953,hastingsMonte1970,morfMonte1986,ciftjaMonte2003} offer a powerful numerical approach in evaluating the integral in polynomial time, by performing importance sampling of the probability density $P(\{z_j\}) = |\Psi_m(\{z_j\})|^2$ and computing expectation values of operators $\hat{O}$ via
\begin{equation}
\langle \hat{O} \rangle = \frac{\int \prod_j d^2z_j \, |\Psi_m(\{z_j\})|^2 \, O(\{z_j\})}{\int \prod_j d^2z_j \, |\Psi_m(\{z_j\})|^2},
\end{equation}
where the integrals are approximated by averages over stochastically generated configurations according to the weight $|\Psi_m(\{z_j\})|^2$.

In practice, a single Monte Carlo step for sampling the Laughlin state consists of proposing a trial move in the configuration space of the particle coordinates $  \{z_j\}  $ and accepting or rejecting it according to the Metropolis-Hastings rule~\cite{metropolisEquation1953,hastingsMonte1970}. Specifically, a randomly chosen particle $  j  $ is displaced by a small complex increment: $z_j \to z_j + \delta z$, where $  \delta z  $ is drawn from a symmetric proposal distribution (typically uniform or Gaussian in the complex plane). The proposed move is then accepted with probability $r = \min\left(1, \frac{P_{\rm new}}{P_{\rm old}}\right)$, where $  P \propto |\Psi_m(\{z_j\})|^2  $ is the Laughlin probability density.

The magnitude of the trial step $  |\delta z|  $ must be carefully tuned: it should be large enough to decorrelate the successive configurations efficiently and allow the walkers to explore the full relevant region of configuration space, yet small enough to avoid very low acceptance rates. When $  |\delta z|  $ is too large, the ratio $  P_{\rm new}/P_{\rm old}  $ typically becomes exponentially small due to the change of energy, leading to very low acceptance and poor sampling efficiency~\cite{morfMonte1986}. As a result, the small step size leads to rapidly growing autocorrelation times with increasing particle number, making Metropolis Monte Carlo less efficient for larger system sizes.

\begin{figure*}[htp!]
\includegraphics[width=\textwidth]{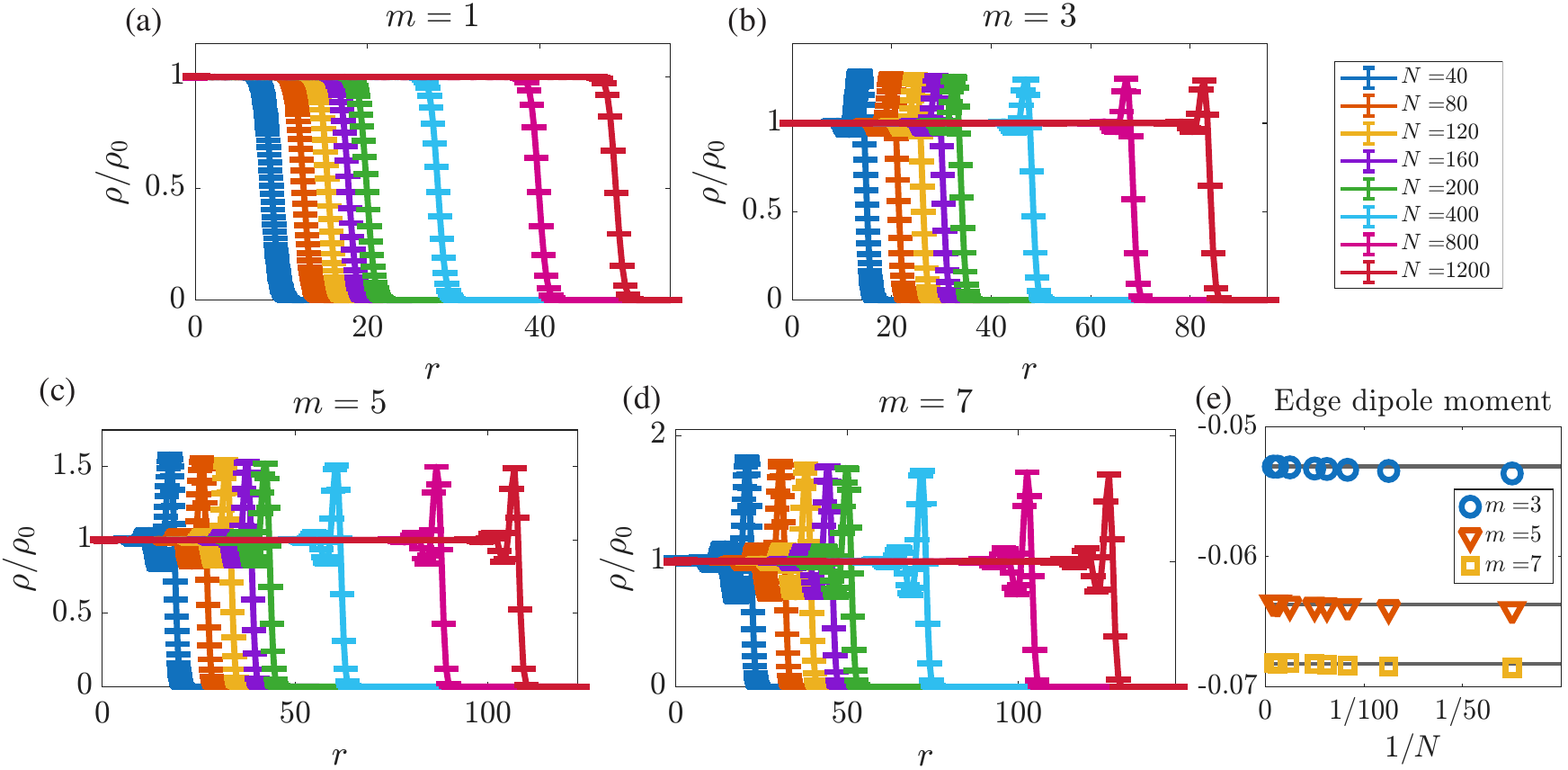}
\caption{\textbf{Electron density of Integer and Laughlin states at different fillings on disk. } (a)-(d) The electron density relative to $\rho_0=\frac{1}{m}$ against the spatial distance to the origin $r$ (in unit of $l_B$) for $m=1,\,3,\,5,\,7$. The four panels share the same legend in the top-right corner, indicating the number of electrons $N$. (e) The edge dipole moment computed from the density profile using Eq.~\eqref{eq:edge_dipole}. The gray horizontal lines are the theoretical prediction $-\frac{1}{4\pi}\frac{m-1}{m}$.}
        \label{fig:rho_Laughlin}
\end{figure*}

On the other hand, hybrid Monte Carlo (HMC) -- a sampling method originated from the high-energy physics community -- offers an appealing alternative~\cite{Scalettar1987,Duane_HMC_1987,Gupta1988,Beyl2018}. Unlike \emph{local} update in Metropolis MC, HMC updates the field configurations \emph{globally} using Hamiltonian dynamics, and is inherently well-suited to parallelization, especially with modern GPU architecture~\cite{fengScalable2025}. However, the spread of HMC from high-energy physics to condensed matter physics communities took a long time. HMC faced difficulties such as poor matrix conditioning and diverging iteration counts in Hubbard-type lattice models~\cite{Scalettar1987,White1988}, which limited its practical advantages. Only more recently, HMC has been successfully applied to spin-fermion and gauge-field coupled to fermions models~\cite{patel2024strange,lunts2023non,fengScalable2025}, where optimizations including preconditioned conjugate gradient solvers, matrix-free multiplications, and GPU acceleration enabled large-scale simulations. With these computational improvements, the non-Fermi-liquid scaling in the fermionic self-energy and bosonic propagators (space-time correlation functions) are observed. More importantly, in the question of U(1) gauge field coupled to fermions in (2+1) dimensions, asymptotic convergence of fermion bilinear correlator and conserved current correlator (unclear previously due to smaller size Metropolish MC simulaion) and their scaling dimensions in good agreement
with field theory are discovered. These developments support the conformal nature of the Dirac spin liquid~\cite{XYXuU12018} and its material realization in quantum magnets~\cite{zengPossible2022,zengSpectral2024,hanSpin2024,BagEvidence24,WuSpinDynamics25,ScheieKYbSe2}.

In light of this background, we now further apply the HMC to FQH systems. Below, we first explain the algorithmic developments in detail and then show the results for both abelian and non-abelian cases.

HMC proposes global updates to all electron positions that respect the probability weight by leveraging energy conservation in classical Hamiltonian dynamics. Recall the plasma analogy, which reinterprets the electron probability density of the Laughlin wave function as a Boltzmann distribution~\cite{halperin1982quantized}:
\begin{equation}
|\Psi_m(\{z_j\})|^2=e^{-V(\{z_j\})},
\end{equation}
where the effective potential
\begin{equation} 
V(\{z_j\}) :=-2m\sum_{i<j}^N \log|z_i - z_j| +\frac{1}{2}\sum_k |z_k|^2
\end{equation}
consists of logarithmic interactions between electrons and a quadratic confining potential. Sampling the Laughlin state is thus equivalent to simulating a classical statistical mechanics model with the partition function
\begin{widetext}
\begin{equation}
Z=\int \prod_j d^2z_j \, e^{-V(\{z_j\})}=\int \prod_j d^2z_j\int \prod_j d^2p_j e^{-V(\{z_j\})-\frac{1}{2}\sum_j|p_j|^2}
\end{equation}
\end{widetext}
where the additional integral over fictitious momenta $p_j$ in the second equality does not affect the distribution as it is independent of $z_j$ and simply factors out as a constant. The system can then be described by the Hamiltonian $\mathcal{H} = V(\{z_j\}) + T(\{p_j\})$, $T(\{p_j\}) = \frac{1}{2} \sum_j |p_j|^2,$
where $p_j$ plays the role of the canonical momentum conjugate to $z_j$. The corresponding Hamiltonian equations of motion are
\begin{equation}
\begin{cases}
    \odiff{z_j}{t}=\pdiff{\mathcal{H}}{p_j}=p_j \\
    \odiff{p_j}{t}=-\pdiff{V}{z_j}=2m\sum_{i\neq j} \frac{z_i-z_j}{|z_i-z_j|^2}-z_i\\
\end{cases}
\label{eq:PDE}
\end{equation}
which conserve the energy $\mathcal{H}$~\footnote{Here, we butcher the notation to express the equations in the same form as real differentials for brevity. The precise definition is to treat the complex variable $z_j$ as two real variables $(x_j,y_j)$, $p_j$ as $(p_{x,j},p_{y,j})$, and take derivatives with respect to each component ($\pdiff{}{z_j} \rightarrow \pdiff{}{x_j}+i\pdiff{}{y_j}=2\pdiff{}{\bar{z}_j}$), which recovers the same right hand sides in Eq.~\eqref{eq:PDE}.}. In principle, evolving the system $(\{z_j\},\{p_j\})$ according to Eq.~\eqref{eq:PDE} for an artificial time $\Delta t$ produces a new configuration $(\{z'_j\},\{p'_j\})$ with $\mathcal{H}' = \mathcal{H}$, yielding a Metropolis acceptance probability of ($r=100\%$). However, in practice, such a PDE cannot be solved analytically and must be integrated numerically using integrators such as the leapfrog algorithm. Since the leapfrog algorithm is a symplectic integrator, although there is numerical error in the time evolution, it can be cured by considering the acceptance ratio $r=e^{\mathcal{H}(\{z_j\},\{p_j\})-\mathcal{H}(\{z'_j\},\{p'_j\})}$, and the overall update still satisfies the detail balance condition. That is, no systematic error is introduced in the process, even when the time evolution is not exact. The error only affects the change in energy and the acceptance ratio as a consequence. In practice, we tune the evolution time $\Delta t$ to control $r$ to around $60\%$. Although in HMC the dimension of parameter space is doubled, i.e., we need to sample both $(\{z_j\},\{p_j\})$, it turns out our HMC update scheme significantly outperforms the usual Metropolis MC in its performance, due to the global nature of the update.

\section{Results}
\label{sec:III}
In this section, we extract universal topological data from large-scale HMC sampling of Laughlin and Moore-Read trial wave functions. First, the topological shifts of the two states are computed, which is a quantized curvature response (Wen-Zee coupling) and is used as a sharp diagnosis of the topological order. 
For the non-Abelian Moore-Read state (given by Pfaffian wave function), we further compute quasihole statistics by evaluating Berry phases and the full non-Abelian Berry matrix under adiabatic exchanges, acting on the topologically degenerate quasihole manifold. 
These results provide the characterization of topological orders and demonstrate the advantages of our HMC method.

\subsection{Laughlin states}
\label{sec:IIIA}
A characteristic manifestation of chiral topological order at fractional filling is the oscillatory structure of the edge density, which forms a universal edge dipole moment per unit length~\cite{Can_Singular_2014, Haldane2009HallViscosity, Read2009NonAbelianAdiabatic, park2014guiding}. 
This dipole can be viewed as the boundary signature of the bulk geometric response (guiding-center spin/Hall viscosity), and controls long-wavelength edge properties in the absence of edge reconstruction~\cite{GromovJensenAbanov2016Boundary}. 

We first computed the electron densities $\rho$ on disk for Laughlin $1/m$ states ($m=1,3,5,7$) as a function of the radial coordinate $r$. It is shown in Fig.~\ref{fig:rho_Laughlin}(a)-(d), reaching system sizes up to $N=1200$ electrons. After normalization by the bulk value $\rho_0=1/m$, the density profiles develop an increasingly wide and flat bulk plateau ($\rho/\rho_0\simeq 1$) as $N$ grows, confirming that finite-size effects are dealt with properly. The integer case $m=1$ exhibits a monotonic depletion at the edge, whereas for fractional Laughlin states ($m=3,5,7$) the density displays pronounced edge oscillations before vanishing. 

The dipole moments can be extracted from the density oscillation and predicted as~\cite{Can_Singular_2014,jiUniversal2024}
\begin{equation}
p_\textrm{edge}=\int_0^\infty \frac{r}{r_0}(r-r_0)\left[m\rho_m(r)-\rho_1(r)\right]dr
=-\frac{1}{4\pi}\frac{m-1}{m},
\label{eq:edge_dipole}
\end{equation}
where $r_0=\sqrt{2mN}$ is the radius of the droplets. 
The subtraction by $\rho_1$ removes the bulk contribution and isolates the universal boundary layer. As we can see from Eq.(\ref{eq:edge_dipole}), $p_\textrm{edge}$ is expected to be topological. The accurate computation of this value, however, is highly challenging, as any small errors in $\rho_m(r)$ will be strongly amplified at large $r$ and $m$. Figure~\ref{fig:rho_Laughlin}(e) shows $p_\textrm{edge}$ extracted from our HMC data for $m=3,5,7$. The computed values converge with increasing $N$ and approach the analytic prediction very well when $N\gtrsim 500$. 
This large-$N$ agreement provides a geometric and topological benchmark for our sampling and confirms that the simulation captures the intrinsic Laughlin edge structure.

\begin{figure}[htp!]
\includegraphics[width=\columnwidth]{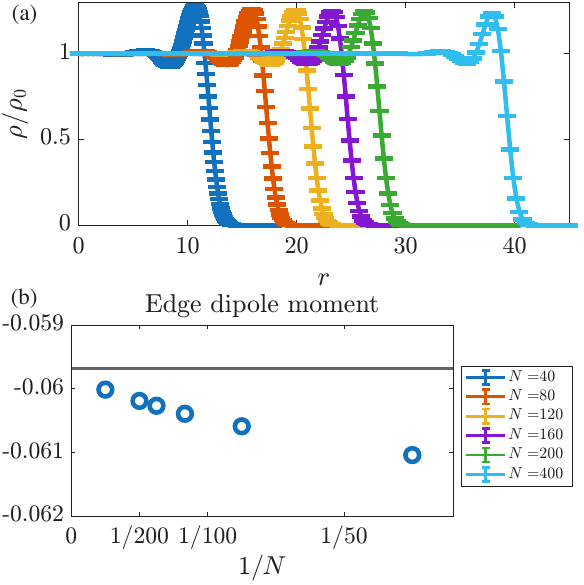}
\caption{\textbf{Electron density of Moore-Read state on disk geometry.} (a) Electron density against the spatial distance to the origin $r$ in a disk geometry, with number of electrons $N=40,80,120,160,200, 400$. (b) The analyzed edge dipole moment for the MR state against $1/N$. The horizontal line is the theoretical prediction $-\frac{3}{16\pi}$ (see main text).}
        \label{fig:rho_MR}
\end{figure}

\subsection{Moore-Read states}
\label{sec:IIIB}
Having tested the validity of our method and computed thermodynamical results of the universal edge structure and dipole moments for Abelian Laughlin phases, we turn to a more intricate FQH phase: the Moore-Read (MR) non-Abelian state, whose trial wave function can be written as a Pfaffian factor multiplying the bosonic Laughlin ($m=2$) Jastrow correlator~\cite{moore1991nonabelions},

\begin{equation}
\Psi_\textrm{MR}(\{z_j\}) = \textrm{Pf}\left(\frac{1}{z_i-z_j}\right)\prod_{i<j}^N (z_i - z_j)^2 \exp\left( -\sum_k |z_k|^2 / 4 \right),
\end{equation}
where $\mathrm{Pf}\!\left(\frac{1}{z_i-z_j}\right)$ denotes the Pfaffian of the $N\times N$ antisymmetric matrix with off-diagonal entries $(i,j)$ being $1/(z_i-z_j)$. 
Compared with the Laughlin sampling, the Pfaffian introduces an additional matrix structure in both the Monte Carlo weight and the HMC ``force''. Evaluating the corresponding updates requires matrix inversion, originating from taking the derivatives of the Pfaffian when solving the equation of motion. This leads to an $O(N^3)$ cost per HMC step, in contrast to the $O(N^2)$ scaling for Laughlin states, though HMC still shows better performance. 

Similar to the Laughlin case in Fig.~\ref{fig:rho_Laughlin} (b)-(d), we computed density oscillations on the edge for the Moore-Read state with $N=40$ to $400$ in Fig.~\ref{fig:rho_MR} (a), from which we can compute the dipole moment as shown in Fig.~\ref{fig:rho_MR} (b). The bulk topological shift $\mathcal{S}$, defined on the sphere by the flux-particle relation $N_\phi=\nu^{-1}N_e-\mathcal{S}$,
measures the coupling of the QH fluid to curvature. For the fermionic Moore-Read Pfaffian at $\nu=\tfrac12$, the shift is $\mathcal{S}=3$, obtained from the spherical Pfaffian ground-state wave function at monopole strength $N_\phi = 2N_e-3$. The mean orbital spin per particle (also called the \emph{guiding-center spin}) $\bar{s}$ is defined by the Wen-Zee curvature coupling and related to the topological shift by $\bar{s}=\mathcal{S}/2$, which is $=3/2$ in this case. Physically, $\bar{s}$ characterizes the intrinsic orbital angular momentum density of the incompressible fluid and controls geometric responses such as the Hall viscosity. It is related to the edge dipole per unit length as $-\frac{\nu\,\bar{s}}{4\pi}=-\frac{3}{16\pi}$, up to the sign convention for the outward normal.

\begin{figure*}[htp!]
\begin{center}
\includegraphics[width=\linewidth]{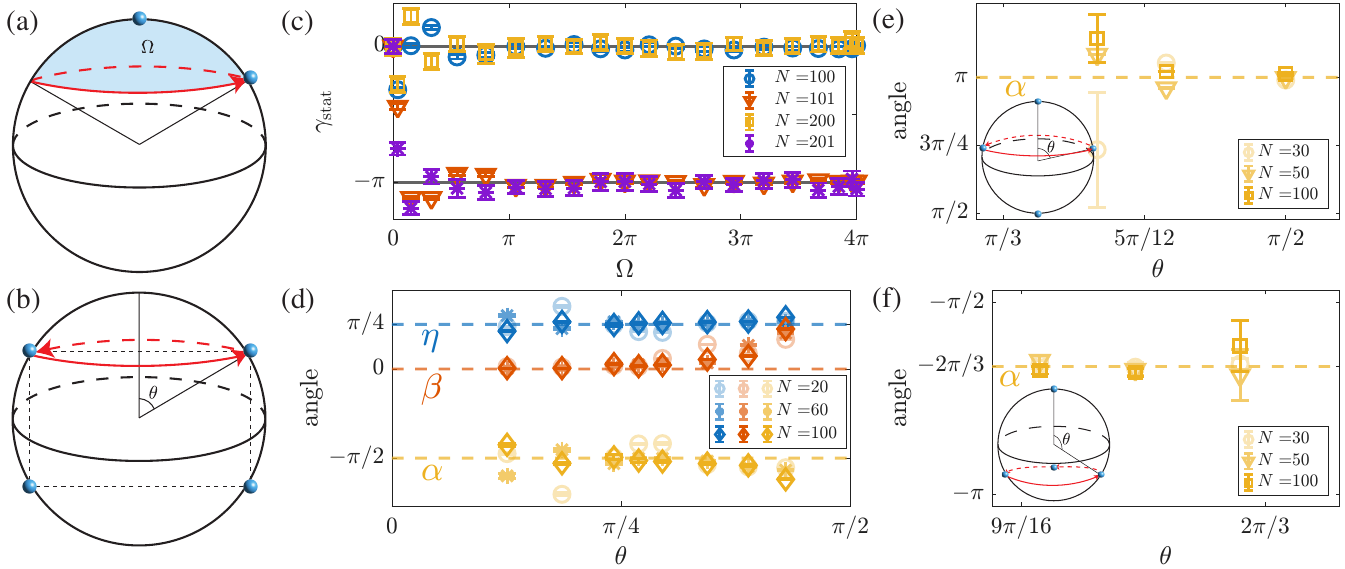}
\caption{\textbf{Berry phase and braiding matrix of Moore-Read state on sphere geometry.} (a) The braiding scheme used to compute the Berry phase of two quasiholes in the Moore-Read state. One electron at a fixed latitude is brought around the other sitting at the north pole, subtending a solid angle $\Omega$ at the center (area of the blue shaded segment). (b) The scheme used to compute the braiding matrix of swapping two quasiholes among 4 quasiholes, each with polar angle $\pm\theta$, forming a tetrahedron. (c) Numerical result of the statistical part of the braiding phase with $N=100,101,200,201$ electrons. The horizontal lines are the theoretical prediction for even and odd sectors, which are 0 and $\pi$, respectively. (d) Result of the braiding matrix using the parameterization in Eq.~\eqref{eq:parameterization}. The blue, orange, and yellow markers denote the numerical results for the angles $\eta$, $\beta$, and $\alpha$, respectively, with $N=20,60,100$ electrons. The dashed lines are $\frac{\pi}{4}$, $0$, and $-\frac{\pi}{2}$ respectively. (e) and (f) shows the $\alpha$ angle for two rotation schemes (shown in insets) comparing to analytic results in Eqs.~\eqref{braiding matrix rotation scheme 1} and \eqref{braiding matrix rotation scheme 2} respectively.}
\label{fig:braiding_result}
\end{center}
\end{figure*}

Beyond the edge dipole moment, the non-Abelian Moore-Read phase supports quasiparticles with nontrivial braiding statistics, making it a prototypical platform for topological quantum computation~\cite{nayak19962n,nayakNonAbelian2008,kitaev2003fault}. To investigate these braiding properties in a setting free of edge effects, we switch to the sphere geometry, where the system is closed, and the bulk topological data (including $\mathcal{S}$) are implemented exactly. To access larger system sizes and improve numerical stability in Monte Carlo sampling, we further apply a double stereographic projection scheme to improve the efficiency of Monte Carlo sampling. The description of spherical geometry and the new projection is detailed in Sec.~\ref{sec:SM_double_projection} of SM~\cite{suppl}. After applying the stereographic projection $z = e^{i\phi}\tan(\theta/2)$ and mapping the particle coordinates back to the complex plane, the structure of the many-body wave function remains unchanged, except for the single-particle form factor. Specifically, the planar Gaussian factor is replaced by $\prod_i \left(\frac{1}{1+|z_i|^2}\right)^{S}$,
where $S=m(N-1)/2$ is the monopole strength, i.e., the total number of flux quanta piercing the sphere.

We first benchmark our method in the case of two quasiholes, where the MR state is still Abelian. As illustrated in Fig.~\ref{fig:braiding_result}(a), we fix one quasihole at the north pole and move the other along a closed loop at constant latitude, enclosing a solid angle $\Omega\in[0,4\pi]$. In this geometry, the Berry phase accumulated along the loop can be written as $\gamma_{\rm Berry}(\Omega)=2\pi\big(\langle L_z\rangle_{\Omega}-\langle L_z\rangle_{\Omega=0}\big)$,
which is the change in the many-body angular momentum expectation value relative to the reference configuration with both quasiholes pinned at the north pole. To isolate the \emph{statistical} contribution, we subtract the $\Omega$-dependent geometric (Aharonov-Bohm) phase. The latter is determined by the fractional charge $e/4$ carried by a Moore-Read quasihole times the magnetic flux threading the enclosed solid angle. Consequently, the statistical part is obtained as $\gamma_{\rm stat}(\Omega)=\gamma_{\rm Berry}(\Omega)-\frac{N}{4}\,\Omega$,
following Refs.~\cite{Prodan_Mapping_2009,trung2023spin}.

The extracted statistical braiding phases are summarized in Fig.~\ref{fig:braiding_result}(b) for systems up to $N=201$ electrons. In the even fusion channel ($N=100$ and $N=200$), $\gamma_{\rm stat}$ shows noticeable fluctuations at small $\Omega$, where the two quasiholes are close and cannot be regarded as well-separated anyons, but rapidly approaches zero as $\Omega$ increases, consistent with bosonic exchange in this channel. In contrast, in the odd fusion channel ($N=101$ and $N=201$), $\gamma_{\rm stat}$ converges to $\pi$ at large $\Omega$, indicating fermionic self-exchange and agreeing with the expected fusion-channel dependence of Moore-Read quasiholes.

Compared to the results in Ref.~\cite{macaluso2019fusion}, where similar measurements were performed with Metropolis MC simulation in disk geometry, our data exhibit clear convergence toward the theoretical prediction for both cases, without the persistent fluctuating tails observed in their radius scaling.
We attribute this improved behavior mainly to the absence of edge effects in spherical geometry. Despite using a comparable number of electrons to theirs, the closed surface eliminates boundary artifacts that plague the disk geometry, leading to cleaner and more reliable anyon statistics.

% \begin{figure}
% \begin{center}
% \includegraphics[width=\linewidth]{sphere_braiding_scheme.pdf}
% \caption{(a) The stereographic projection maps a braiding scheme on the sphere to the equivalent process on the disk. (b) The braiding scheme involving four quasiholes used in the main text and the equivalent braiding on the disk. (c) Tree diagram corresponding to the creation and braiding of four Moore-Read quasiholes with the braiding scheme in (b).}
% \label{fig:sphere_braiding}
% \end{center}
% \end{figure}

\subsection{Non-Abelian Braiding}
\label{sec:IIIC}
Next, we discuss the braiding matrix for the MR state if we introduce four quasiholes with charge $e/4$ into the system and braid them. In this case, specifying the positions of four quasiholes results in two degenerate states given by the first-quantized wave functions of the form
\begin{widetext}
    \begin{equation}
    \label{MR 2qh state}
    \Psi_{\text{MR}}(\{z_i\};\{\eta_j\})=\text{Pf}\left[\frac{(z_i-\eta_1)(z_i-\eta_2)(z_j-\eta_3)(z_j-\eta_4)+(i\leftrightarrow j)}{z_i-z_j}\right]\prod_{i<j}(z_i-z_j)^2\exp\left({\sum_{i}|z_i|^2/4}\right)
\end{equation}
\end{widetext}
where $\{\eta_i\}=(\eta_1,\eta_2,\eta_3,\eta_4)$ are the positions of the four quasiholes and ``$(i\leftrightarrow j)$" denotes the term obtained by swapping the $i$ and $j$ indices. Using the same form of \Cref{MR 2qh state}, different wave functions can be constructed by permuting the order of $\{\eta_1,\eta_2,\eta_3,\eta_4\}$, resulting in different wave functions describing quasiholes at the same four locations. These wave functions span a Hilbert space of dimension 2~\cite{moore1991nonabelions,nayak19962n,tong2016lectures}, and as a result, the outcome of the braiding process is described by a 2-by-2 unitary matrix~\cite{suppl}. For the purpose of the later discussion, here we briefly revise the numerical method for computing this matrix, following Ref.~\cite{tserkovnyak2003monte}. We start by considering a braiding scheme in which one or more quasiholes are moved along some trajectory parametrized by a general vector $\mathbf R$ in configuration space, which depends on some "fictitious" time parameter $\mathbf R = \mathbf R(t)$. We discretize this path into $N_s$ steps: $\mathbf R_n\equiv \mathbf R(t_n)$, for $t_0<t_1<...<t_N$. We impose that the state must come back to its initial configuration: $\mathbf R(t_N)=\mathbf R(t_0)$. The unitary matrix describing the dynamics of the whole process can then be computed recursively as%such a process will introduce a braiding matrix as detailed in Sec.~\ref{sec:SMII} in SM~\cite{suppl} and the corresponding numerical scheme in obtaining such braiding matrix is detailed in Sec.~\ref{sec:SM_braiding_mat} in SM~\cite{suppl}. 
\begin{align}
    U(t_{n+1}) &= U(t_n)\left(1+\frac12 A(t_n)\right)\left(1-\frac12 A(t_n)\right)^{-1}\label{recursive1}\\
    U(t_0)&=\begin{pmatrix}1&0\\0&1\end{pmatrix}\label{recursive2}
\end{align}
where $A(t_n)$ is a $2\times2$ matrix whose entries are given by
\begin{align}
\label{A matrix}
A(t_n)_{ab}=\langle\psi_a(t_n)|\psi_b(t_{n+1})\rangle - h.c.
\end{align}
Here $|\psi_1(t)\rangle$ and $|\psi_2(t)\rangle$ is the basis for the Hilbert space at time $t$. Any choice of basis can be used for this computation so long as it is orthonormal and continuous in $t$. For our calculation, we choose the following basis~\cite{suppl}:
\begin{align}
    |\psi_1(t)\rangle &=\frac{1}{\sqrt{2+2|o|}}\left (\ket{\psi_{(1,2,3,4)}}+\frac{o^*}{|o|}\ket{\psi_{(1,3,2,4)}} \right ), \label{basis1}\\
    |\psi_2(t)\rangle &=\frac{1}{\sqrt{2-2|o|}}\left (\ket{\psi_{(1,2,3,4)}}-\frac{o^*}{|o|}\ket{\psi_{(1,3,2,4)}} \right ), \label{basis2}
\end{align}
where $\ket{\psi_{(\eta_1,\eta_2,\eta_3,\eta_4)}}\sim\psi_{\text{MR}}(\{z_i\};(\eta_1,\eta_2,\eta_3,\eta_4))$ denotes the state whose first-quantized wave functions are given by Eq.\ref{MR 2qh state} and $o=\overlap{\psi_{(1,2,3,4)}}{\psi_{(1,3,2,4)}}$. Note that these two states are always orthonormal by design. Using these basis states, the matrix $A$ in Eq.~\eqref{A matrix} is computed by the HMC method.
% On the sphere, the trajectory of a particle must be regarded as an equivalent motion on the disk via the stereographic projection (see Fig.~\ref{fig:sphere_braiding}). In particular, since the south pole of the sphere corresponds to the points at infinity, while a particle at the south pole seemingly remains stationary as the sphere rotates, it actually corresponds to a particle moving along a circle of infinite radius on the disk. By carefully tracing the position of each quasihole on the sphere, the exact braiding scheme can be determined.

% In Fig.~\ref{fig:braiding_result} (c), we show the braiding that is equivalent to the quasihole arrangement in the four-quasihole calculation used in the main text, and Fig.~\ref{fig:braiding_result} (c) shows the corresponding tree diagram. By carefully resolving each crossing along the tree diagram, one can calculate that the total Berry matrix of this process is given by
% \begin{equation}
% \label{Berry matrix outcome}
% \mathbf U = e^{i\gamma} e^{-i\pi/4}
% \begin{pmatrix}
% 1 && 0 \\ 0 && 1
% \end{pmatrix}
% \end{equation}
% where $\gamma$ is some $\Omega$-dependent scalar phase.

Following the braiding scheme in Ref.~\cite{tserkovnyak2003monte}, two quasiholes are adiabatically exchanged while the other two remain fixed, as shown in Fig.~\ref{fig:braiding_result}(c). The braiding contour is discretized into $N_s=320$ segments, and the unitary braiding matrix is accumulated along the path. We used the same parameterization of the matrix as that in Ref.~\cite{tserkovnyak2003monte}, that is,
\begin{equation}
U = e^{i\chi} 
\begin{pmatrix}
e^{i\eta}\cos{\beta/2} && ie^{-i\epsilon/2}\sin{\beta/2} \\
ie^{i\epsilon/2}\sin{\beta/2} && e^{-i\eta}\cos{\beta/2}
\end{pmatrix}.
\label{eq:parameterization}
\end{equation}

Note that the actual outcome of the braiding process (i.e. the matrix $U$ computed in our numerics) may differ from the theoretically predicted value due to different choices of basis (it is known that braiding matrices are not basis-independent), and an additional scalar phase factor coming from the Aharonov-Bohm phase and the parallel transport of the quasihole spin along the curvature of the sphere. However, the \emph{ratio between eigenvalues} of the matrix is invariant with respect to both of these factors. Since the matrix is unitary, the ratio can be represented as a phase $e^{i\alpha}$, parametrized by angle $\alpha$. The three angle parameters $\eta$, $\beta$, and $\alpha$ serve as a reliable way to compare numerical results with theoretical calculations.

The numerical result is shown in Fig.~\ref{fig:braiding_result} (d). When the anyons are separated far from each other ($\theta\approx\pi/4$), the angles $\eta$ (the half of relative phase between diagonal elements) and $\beta$ (magnitude of off-diagonal elements) flatten at $\pi/4$ and $0$ respectively, giving the braiding matrix 

\begin{equation}
U_{12} \approx e^{i\chi} 
\begin{pmatrix}
1 && 0 \\ 0 && -i
\end{pmatrix}.
\label{eq:braiding_matrix_result}
\end{equation}
This result agrees well with the single-exchange braiding matrix~\cite{Bonderson2011,tong2016lectures,suppl}. We found that the convergence as a function of $N=20,60,100$ is most obvious around $\theta\sim\pi/4$, as this is the angle at which all four anyons are the most well-separated on the sphere. Overall, our convergence is much better than that in Ref.~\cite{tserkovnyak2003monte} with local updates of the conventional Metropolis MC scheme; we attribute this improvement to the more efficient global update in our HMC, which enables it to reach the thermodynamic limit faster and more reliably.

Furthermore, our choice of orthonormal basis in Eqs.~\eqref{basis1} and \eqref{basis2} %, detailed in Sec.~\ref{sec:SM_braiding_mat} of the Supplemental Material~\cite{suppl}, 
 yields the nearly diagonal braiding matrix directly, up to an overall global phase $\chi$, without requiring additional basis transformations. This outcome is highly nontrivial, given that the braiding matrix is inherently basis-dependent and related to other representations via unitary transformations.

We further propose two alternative braiding schemes and compute the corresponding braiding matrix using our method. Similar to what is done for extracting the Berry phase, the advantage of these schemes is that the braiding path is generated by rotation of the entire sphere around the $z$-axis. Thus, by making use of the rotational symmetry of the sphere, we can reduce the computation of wave function overlaps at $N_s$ steps into just one step:
\begin{equation}
    U(t_{N_s}) = \left[\left(1+\frac12 A(t_n)\right)\left(1-\frac12 A(t_n)\right)^{-1}\right]^{N_s}
\end{equation}

The schematic of these two new braiding schemes is shown in the insets of Fig.~\ref{fig:braiding_result} (e) and (f). In the first scheme, two quasiholes are placed at opposite poles, and the remaining two are placed across each other on a circle of latitude. The positions of these two quasiholes are exchanged when the sphere rotates by 180$^\circ$ around the $z$-axis. This results in a sequence of braids that accumulates into a braiding matrix given, up to a scalar factor and a basis transformation ~\cite{suppl}, by:
\begin{equation}
    \label{braiding matrix rotation scheme 1}
    B=\begin{pmatrix}-1&0\\0&1\end{pmatrix}
\end{equation}

In the second rotational braiding scheme, one quasihole is placed at the north pole while the remaining three are placed at the vertices of an equilateral triangle inscribed within a circle of latitude. The four quasiholes form the vertices of a tetrahedron, which rotates into itself when the sphere rotates by 120$^\circ$. In this case, the quasihole exchange is more complicated, but the process is still a closed loop in parameter space since all quasiholes are identical. The outcome of this process is given by the braiding matrix:
\begin{equation}
    \label{braiding matrix rotation scheme 2}
    B=\frac{1}{\sqrt{2}}\begin{pmatrix}i&1\\i&-1\end{pmatrix}
\end{equation}

Note that of the three angle parameters investigated previously, only $\alpha$ is both gauge-independent and basis-independent. Hence, it is the only relevant quantity that can be measured in an experiment. Thus, we use this quantity to compare the numerical results with Eq.~\eqref{braiding matrix rotation scheme 1}. For these two cases, we choose $N_s=1200$ to reduce the discretization error. As shown in Fig.~\ref{fig:braiding_result} (e), the ratio between the two eigenvalues is around $\pi$, which is consistent with Eq.~\eqref{braiding matrix rotation scheme 1}. Similarly, the ratio between the eigenvalues of Eq.~\eqref{braiding matrix rotation scheme 2} is $e^{-i2\pi/3}$. This is also consistent with the numerical result, as can be seen in Fig.~\ref{fig:braiding_result}(f). To the best of our knowledge, this is the first computation of the braiding matrix for a non-Abelian braiding process involving more than one pairwise exchange. 

\section{Discussions}
\label{sec:IV}
As shown above, our hybrid Monte Carlo method can efficiently compute physical observables from samplings of the Laughlin and Moore-Read wave functions in fractional quantum Hall systems. With computational advancements, including global update and double stereographic projection on spherical geometry, our hybrid Monte Carlo simulation is significantly faster than the widely used Metropolis Monte Carlo scheme. 

Recently, due to the rapid developments in fractional quantum anomalous Hall (FQAH) states~\cite{Regnault2011_FCI,wuAdiabatic2012,Roy2014_band_geometry_fci,Wang2021_geometry_flatband, Ledwith2023_vortexability,Hongyu_prl2024_thermodinamicFCI,luContinuous2025,Lu2024Interaction}, the relation between the FQAH and FQH wave functions is becoming a central topic both theoretically and experimentally~\cite{Cai2023_signature_fqah_mote2,Park2023_observation_fqah_mote2,Zeng2023_thermo_evidence_fqah_mote2,Xu2023_Observation_FQAH_tMote2,Lu2024_FQAH_multilayer_graphene}. In Landau levels with uniform quantum geometric tensor (QGT), anyons described by FQH wave functions can form ``molecule-like" bound states due to effectively attractive interaction between them (despite the repulsive interaction between electrons)~\cite{xuDynamics2025, gattu2025molecular, liu2025characterization}. The braiding of these fractionally charged anyons can also be affected by additional nonuniform Berry curvature, even with a uniform magnetic field, because of their shape deformation in real space~\cite{trung2023spin}. For topological phases realized in new experimental platforms without an external magnetic field, both the dynamics and braiding of the anyons can be strongly affected by the nontrivial fluctuation of the QGT both in real and momentum space. We expect rich geometric properties in generic Chern bands to strongly affect the physical measurement and manipulation of anyons. However, comparing and contrasting the anyonic properties in the LLs and generic Chern bands would require extensive numerical computations. The HMC method we developed can potentially be very useful. This is especially true in the context of ideal Chern bands which are good approximation to a number of experimentally realized quantum materials, due to the close analogy between FQH wave functions and FQAH wave functions in the ideal Chern band.

In addition, one experimental approach to realize an exact ideal Chern band is to use massless Dirac fermions placed in an inhomogeneous magnetic field~\cite{aharonovGround1979}. In such systems, the question of whether the fractional-filled ideal Chern band system still forms the FQH liquid has been raised, and the possibility that the FQH liquids (both Laughlin and Moore-Read) are not stable in an inhomogeneous magnetic field towards a gapless dielectric state through Berezinskii-Kosterlitz-Thouless (BKT) transition~\cite{moitraInstability2025}, even for the filling of $1/3$, has been pointed out. 

In the technically similar but conceptually different direction of topological quantum computation, the FQH states (both Laughlin and Moore-Read) have also been argued to be unstable 
under density decoherence~\cite{wangFractional2025} -- to mimic coupling of local charge density to non-thermal noise in the quantum circuit settings. It is analytically argued that there exists a critical filling factor $\nu_c$, above which the quantum information remains fully recoverable for arbitrarily strong decoherence, and $\nu = 1/3$ Laughlin state and $\nu = 1/2$ MR state both lie within this range. Below $\nu_c$, both classes of states are expected to undergo a decoherence-induced BKT transition into a critical decohered phase. It is further argued that for Laughlin states, information encoded in the topological ground state manifold degrades continuously with decoherence strength inside this critical phase, vanishing in the limit of infinite decoherence strength, but the quantum information encoded
in the fusion space of non-abelian anyons of the MR states remains fully recoverable for arbitrary filling.

These interesting theoretical speculations on the instability of FQH states, induced by either an inhomogeneous magnetic field or quantum decoherence, deserve systematic, unbiased numerical investigations, and these simulations necessarily require state-of-the-art numerical methods. In particular, one foresees that the power-law correlation in the proposed BKT transitions requires extremely large system sizes and a stable simulation technique to truly understand the fate of both Abelian and, more importantly, non-Abelian anyons under inhomogeneous magnetic field and quantum decoherence. With the thermodynamic limit results obtained at ease, we anticipate that our hybrid Monte Carlo scheme will enable the assessment of the necessary system sizes that can unambiguously verify the associated essential singularity and power-law correlations with changing exponents~\cite{jiangSolving2022,jiangMonte2022}. In this way, our HMC simulation technique will be of widespread usage in the active research direction of understanding the FQH and FQAH wave functions and the topological
order in mixed quantum states~\cite{chenSymmetry2024,wangAnalog2025,SangStability2025}. 

Other interesting applications of our HMC method include extracting the long-wavelength limit of the static structure factor for both Laughlin and Moore-Read states to match the predictions for the coefficients such as that of the $q^6$ term~\cite{dwivediGeometri2019}; further speeding up the Monte Carlo sampling for wave functions requiring (anti)symmetrization~\cite{boseMonteCarlo2025}; calculating correlation function as well as verifying Ward identities related to the perturbation of the background magnetic field and background curvature of FQH states in curved space~\cite{canFractional2014,canGeometry2015,nguyenParticle2017}; obtaining the "quantization" of the Hall viscosity in terms of the Berry curvature of adiabatically deforming the FQH states on torus geometry~\cite{avronViscosity1995,fremlingHall2014}, etc.  All these questions require a simulation technique that can handle large sizes with better quality data, and we believe our HMC, with its global update and faster simulation speed, will become the method of choice to address them.

\begin{acknowledgments}
We thank Cenke Xu, Marcello Dalmonte, and Inti Sodemann for inspiring discussions on the stability of FQH wave functions. We thank Ajit C. Balram, Dung Xuan Nguyen, and Leonardo Mazza for the constructive comments on the manuscript and suggestions of references. TTW, ML, and ZYM acknowledge the support from the Research Grants Council (RGC) of Hong Kong (Project Nos. 17309822, C7037-22GF, 17302223, 17301924, 17301725), the ANR/RGC Joint Research Scheme sponsored by RGC of Hong Kong and the French National Research Agency (Project No. A\_HKU703/22). 
We thank the HPC2021 system under the Information Technology Services at the University of Hong Kong~\cite{hpc2021}, as well as the Beijing Paratera Tech Corp., Ltd~\cite{paratera} for providing HPC resources that have contributed to the research results reported within this paper. This work at Nanyang Technological University, Singapore, is supported by the National Research Foundation, Singapore, under the NRF Fellowship Award (NRF-NRFF12-2020-005), Singapore Ministry of Education (MOE) Academic Research Fund Tier 3 Grant (No. MOE-MOET32023-0003) “Quantum Geometric Advantage”, and Singapore Ministry of Education (MOE) Academic Research Fund Tier 2 Grant (No. MOE-T2EP50124-0017).
\end{acknowledgments}

\bibliographystyle{longapsrev4-2}
\bibliography{ref}

%%%%%%%
%
% 			SUPPLEMENTARY GOES HERE
%
%%%%%%%
\clearpage

\onecolumngrid

\renewcommand{\thesection}{S\arabic{section}}
\renewcommand{\thefigure}{S\arabic{figure}}
\renewcommand{\theequation}{S\arabic{equation}}
\renewcommand{\thepage}{S\arabic{page}}
\setcounter{figure}{0}
\setcounter{page}{0}
\setcounter{section}{0}

\begin{center}
{\large \textbf{Supplementary Material for “Hybrid Monte Carlo for Fractional Quantum Hall States”}}

\vspace{1cm}

\noindent\mbox{%
    \parbox{\textwidth}{%
        \indent 
In this Supplementary Material, we introduce the basic notation of the Berry matrix (Sec.~\ref{sec:SMI}) and apply it to the analysis of fusion and braiding in anyon models of FQH systems (Sec.~\ref{sec:SMII}). We also explain our invention of the double stereographic projection on the sphere such that the global update of the HMC can be made to perform at its best (Sec.~\ref{sec:SM_double_projection}) and the numerical scheme of obtaining the braiding matrix in the MR case with 4 anyons (Sec.~\ref{sec:SM_braiding_mat}).
    }%
}
\end{center}

\section{Berry matrix}
\label{sec:SMI}
%\subsection{Derivation of the Berry matrix}
The Berry matrix is the generalization of the Berry phase for an adiabatic process involving degeneracy. In this section, we provide a formal definition in terms of adiabatic transport. Suppose we have a Hamiltonian $\hat H(\mathbf R(t))$ which varies as a function of some parameters $\mathbf R$, which in turn varies in time $t$. We assume that at any time $t$, the ground state is $n$-fold degenerate:
\begin{equation}
\label{ground state}
\hat H(\mathbf R(t))|\psi_{\alpha}(\mathbf R(t))\rangle = E(\mathbf R(t))|\psi_{\alpha}(\mathbf R(t))\rangle, \alpha=1,2,...,n
\end{equation}
where $\{|\psi_{\alpha}\rangle,\alpha=1,2,...,n\}$ forms an orthonormal basis: $\langle\psi_\alpha|\psi_\beta\rangle = \delta_{\alpha\beta}$. We consider the dynamic of any state within this degenerate ground state:
\begin{equation}
\label{state}
|\psi(\mathbf R(t))\rangle = \sum_{\alpha}c_\alpha|\psi_\alpha(\mathbf R(t))\rangle
\end{equation}
for some complex coefficients $c_1,c_2,...,c_n$ satisfying $\sum_{\alpha=1}^{n}|c_\alpha|^2=1$. The time-evolution of this state must obey the Sch\"odinger equation:
\begin{equation}
\label{Schrodinger}
i\frac{d}{dt}|\psi(t)\rangle = \hat H|\psi(t)\rangle
\end{equation}
(Note that from here to keep the notations clean, we consider time $t$ as a parameter and skip the implicit $\mathbf R$-dependence of the state. Since the state must be normalized at all time, we can describe the evolutions of the coefficients $c_\alpha(t)$ by a unitary matrix:
\begin{equation}
\label{unitary matrix}
c_\alpha(t) = \mathcal U_{\alpha\beta}c_{\beta}(0)
\end{equation}
Here to keep the notations clean but clear, we will use Einstein's summation convention only for the matrices and scalar coefficients (sum over repeated indices \emph{regardless whether they are upper or lower indices}). Summation convention is \emph{not} used if the repeated indices involve a bra- or ket-vector, in which case an explicit sum must be written such as in \Cref{state}. Using the matrix $\mathbf{\mathcal U}(t)$, we can write the state in \Cref{state} and its time-derivative as
\begin{align}
|\psi(t)\rangle &= \sum_\alpha \mathcal U_{\alpha\beta}c_\beta|\psi_\alpha(t)\rangle\\
|\dot\psi(t)\rangle &= \sum_\alpha \left[\dot{\mathcal U}_{\alpha\beta}c_\beta(0)|\psi_\alpha(t)\rangle +\mathcal U_{\alpha\beta}c_\beta(0)|\dot\psi_\alpha(t)\rangle\right]
\end{align}
Substituting these into the Schr\"odinger equation gives the matrix equation:
\begin{equation}
\label{matrix eqn}
\dot{\mathbf{\mathcal{U}}}\cdot\mathbf c = iE\mathbf{\mathcal U}\cdot \mathbf c - i\mathbf{\mathcal U}\cdot\mathbf{A}\cdot\mathbf c
\end{equation}
where $\mathbf c\equiv(c_1(0),c_2(0),...,c_n(0))^{\text{T}}$ and $\mathbf{A}$ is a matrix whose element is given by
\begin{equation}
\label{Berry connection}
A_{\alpha\beta} = i\langle\psi_\alpha(t)|\dot\psi_\beta(t)\rangle
\end{equation}
The solution to \Cref{matrix eqn} is
\begin{equation}
\label{matrix solution}
\mathbf{\mathcal U}(t)=e^{i\int_0^t\mathbf{A}(t')dt'}e^{-i\int_0^tE(t')dt'}
\end{equation}
When the state moves adiabatically in a closed loop in parameter space such that $\mathbf R(T)=\mathbf R(0)$ for very large $T$, the second term in \Cref{matrix solution} can be made trivial by adding a constant term to the overall Hamiltonian, and we find the effect of this transformation is a $U(n)$ phase given by
\begin{equation}
\label{Berry matrix}
\mathbf{\mathcal U} = \exp\left({-\oint\langle\psi_\alpha|\nabla_\mathbf{R}\psi_\beta\rangle\cdot d\mathbf R}\right)
\end{equation}
which is the definition of the Berry matrix.

\section{Fusion and Braiding in Anyon Models}
\label{sec:SMII}
\subsection{Fusion product and fusion rule}
An anyon model is specified by a set of anyon species (a.k.a. ``topological charges'') $a,b,c,...$ along with a set of \emph{fusion rule} of the form
\begin{equation}
\label{fusion}
a\times b=\sum_c N_{ab}^c c
\end{equation}
where $\times$ denotes the fusion product and $N_{ab}^c$ is a non-negative integer specifying how many ways $a$ and $b$ can be fused into $c$. In the present case of interest, it is enough to assume that $N_{ab}^c$ can only be either 0 ($a$ and $b$ do not fuse into $c$) or 1 ($a$ and $b$ fuse into $c$ by a unique channel). In every anyonic model there is also a ``vacuum'', which is denoted by the symbol $1$ and has the trivial fusion rule:
\begin{equation}
\label{anyon vacuum}
1\times a = a\times 1 = a
\end{equation}

The reverse of a fusion process is called \emph{splitting}. A state consisting of anyon species $a$ and $b$, splitted from an anyon $c$, can be represented by a \emph{tree diagram}
\begin{equation}
\label{abc}
|a,b;c\rangle \sim \includegraphics[width=30pt]{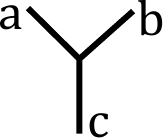}
\end{equation}
The tree diagram can be thought of as a ``Feynmann-diagram-type'' process with time going upward. The use of the ket notation in \Cref{abc} implies that these states form an inner product space with well-defined vector addition, scalar multiplication, and inner product. However, we will skip this discussion for now because it is not necessary for the focus of the discussion below, which is deriving the braiding matrix of any braiding process in a given anyon model. An extended introduction to the fusion and splitting vector spaces can be found in e.g., Ref. ~\cite{bonderson2008interferometry}.

\subsection{Fusion Matrices}
A state consisting of $n$ anyons $a_1$, $a_2$,..., $a_n$ can be represented by a series of splitting from vacuum\footnote{In these discussion we almost always use vacuum as the starting point, but that is not necessarily the case even in a physical system. For example, in the odd-parity sector of the Moore-Read state, the starting point should be a single Majorana fermion.}
\begin{equation}
\label{a1a2..an}
\includegraphics[width=0.2\linewidth]{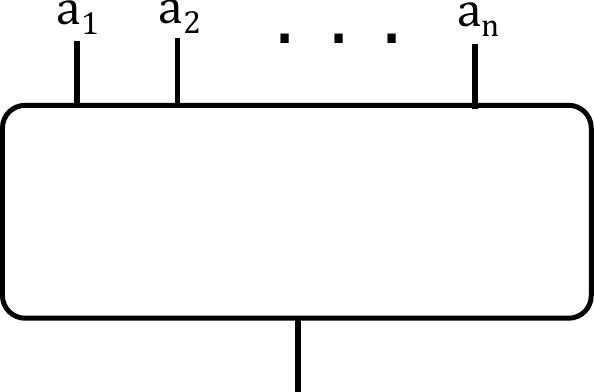}
\end{equation}
where the dashed outline box represents a series of splittings that finally result in the anyons $a_1$, $a_2$,...$a_n$ on top. Since anyons can only be created in pairs, we can see that there are multiple ways to fill in the processes in the box. For example, for three anyons, we have two possible splitting processes:
% \begin{align}
% \includegraphics[width=0.1\linewidth]{abcm}\\
% \includegraphics[width=0.1\linewidth]{abcn}
% \end{align}
where $m$ and $n$ denote some intermediate anyon species. These two states are related by a \emph{fusion matrix} $F_{abc}^1$ defined as
\begin{equation}
\label{fusion matrix}
\includegraphics[width=0.3\linewidth]{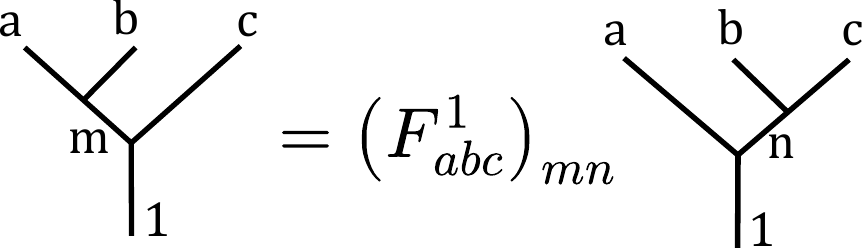}
\end{equation}
Here, an implicit sum over the index $n$ on the RHS is used. For a state with four or more anyons, there can be many more splitting processes that have the same outcomes. Consider the following two diagrams:
\begin{equation}
\includegraphics[width=0.11\linewidth]{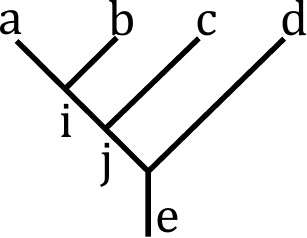}
\text{ and }
\includegraphics[width=0.11\linewidth]{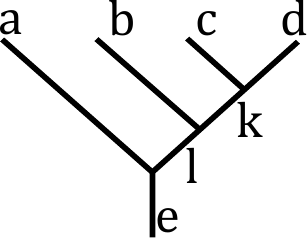}
\end{equation}
To relate from one state to the other, we repeatedly pick a subtree of one and apply on it the fusion matrix as per Eq.~\eqref{fusion matrix}. We can see that by manipulating the diagrams this way, there are two ways to get from one to the other:
\begin{equation}
\label{pentagon}
\includegraphics[width=0.6\linewidth]{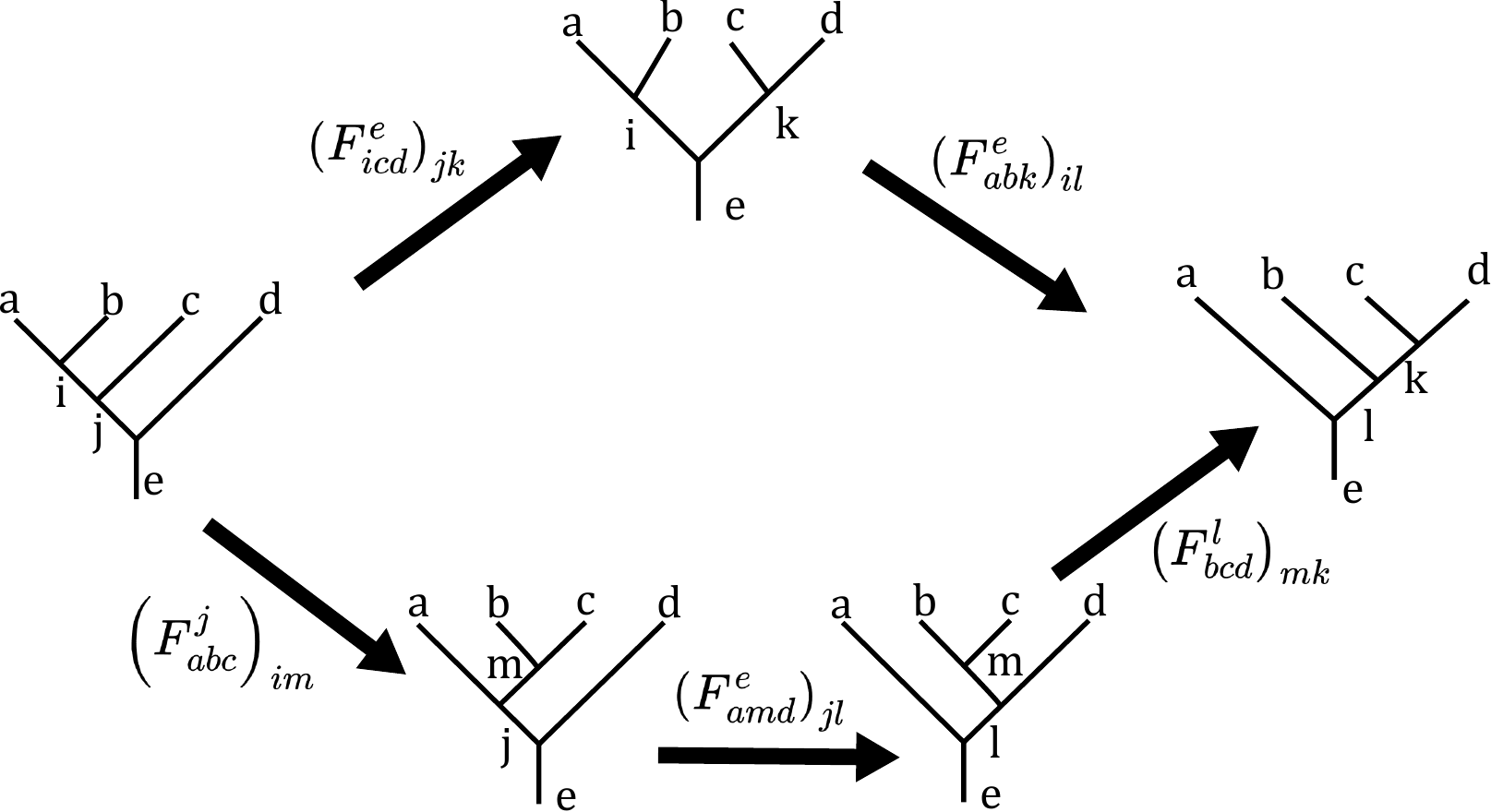}
\end{equation}
Thus, to ensure consistency at the final process, the fusion matrices must follow the \emph{pentagon rule}:
\begin{equation}
\label{pentagon rule}
\left(F_{abk}^{e}\right)_{il}\left(F_{icd}^{e}\right)_{jk}=\sum_m\left(F_{bcd}^{l}\right)_{mk}\left(F_{amd}^{e}\right)_{jl}\left(F_{abc}^j\right)_{im}
\end{equation}
It turns out that this pentagon rule is sufficient to ensure consistency for every splitting process.

\subsection{Braiding Matrices}
Note that the fusion matrices allow us to relate different splitting processes only if the order of the anyons in the final outcome is the same. If two processes result in the same anyon species but in different orders, they can be related by a braiding process, described by the \emph{R-matrix} as
\begin{equation}
\label{eq:R-matrix}
\includegraphics[width=0.17\linewidth]{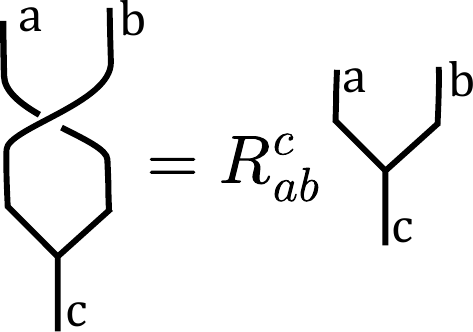}
\end{equation}
For multiple anyon states, there may be multiple ways to braid pairs of anyons to go from one tree diagram to another. Similar to the pentagon rule for the fusion matrices, the R-matrices must also follow another consistency rule called the \emph{hexagon rule}, which ensures that the following diagram commutes:
\begin{equation}
\includegraphics[width=0.63\linewidth]{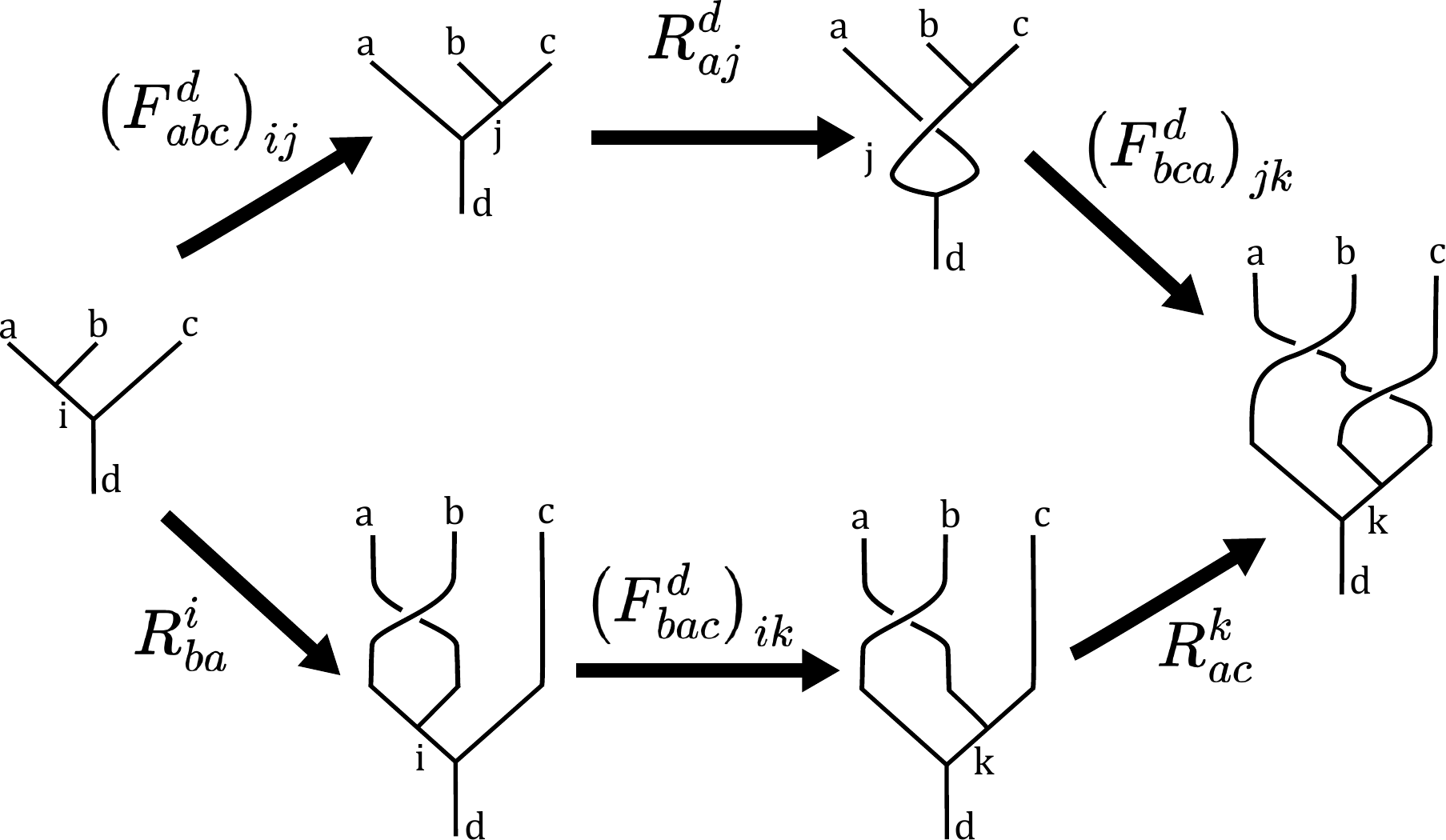}
\end{equation}

\subsection{Fusion and Braiding Matrices in the Ising Anyon Model}
The Ising anyon model contains three anyon species, denoted by $1$ (i.e., the vacuum), $\sigma$, and $\psi$. They obey the following set of fusion rules:
\begin{align}
\sigma\times\sigma&=1+\psi\label{Ising1}\\
\psi\times\psi&=1\label{Ising2}\\
\psi\times\sigma&=\sigma\label{Ising3}
\end{align}
The fusion matrices and R-matrices in this model are as follows ~\cite{kitaev2006anyons}
\begin{align}
R_{\psi\psi}^1&=-1\\
R_{\sigma\sigma}^1&=e^{-i\pi/8}\\
R_{\sigma\psi}^{\sigma}&= -i\\
R_{\sigma\sigma}^{\psi}&=e^{i3\pi/8}\\
F_{\sigma\sigma\sigma}^\sigma &= \frac{1}{\sqrt{2}}
\begin{pmatrix}
1 && 1 \\ 1 && -1
\end{pmatrix}
\end{align}
In the Moore-Read state, the $\sigma$-anyon is a single charge-$e/4$ quasihole, while the $1$-anyon and $\psi$-anyon are two types of charge-$e/2$ quasiholes. The former is a single magnetic flux, and the latter is a magnetic flux dressed by a Majorana fermion. A four-quasihole~\footnote{In this text, a ``quasihole'' refers simply to a charge-$e/4$ quasihole, i.e., a $\sigma$-anyon.} state can be represented by a tree diagram:
\begin{equation}
\label{eq:four qh tree}
\includegraphics[width=0.18\linewidth]{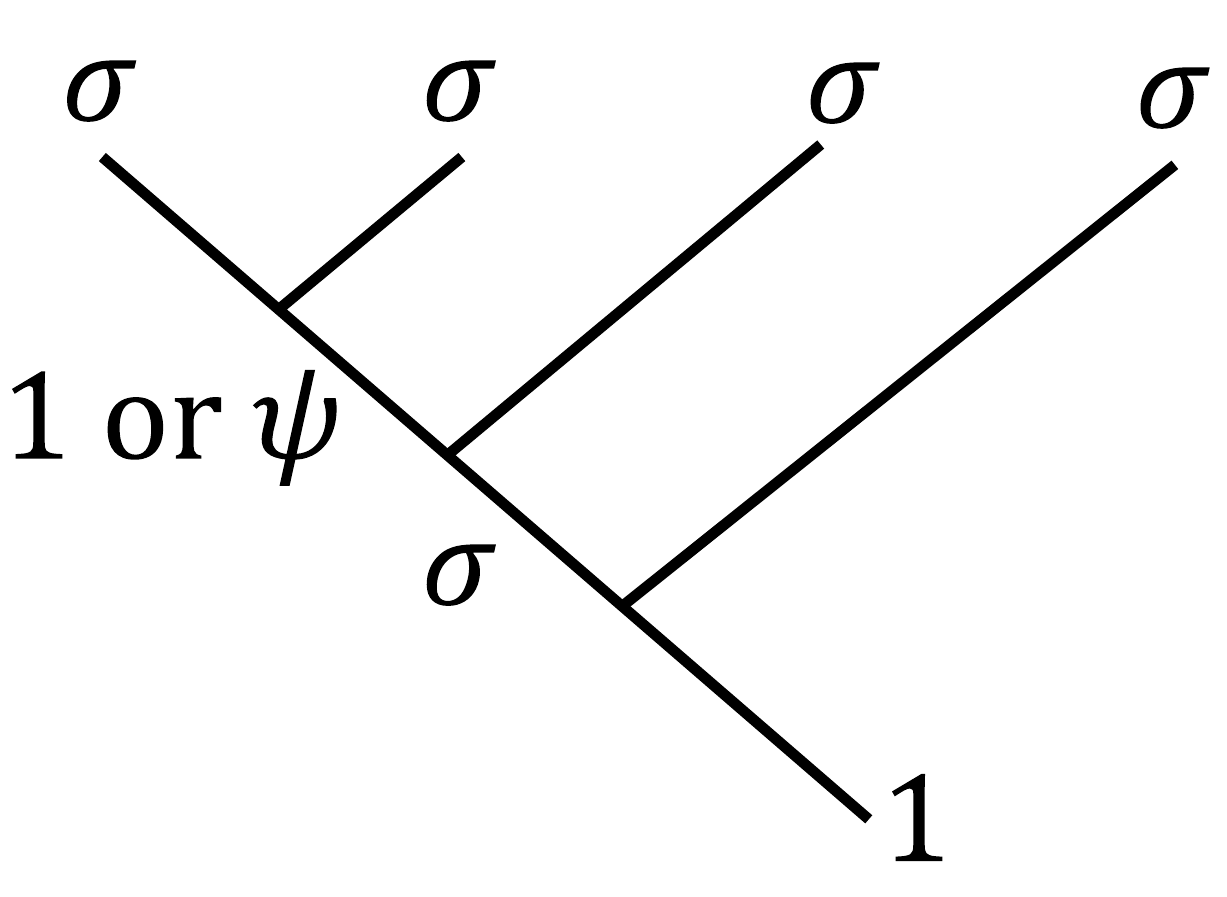}
\end{equation}
There are two such states, each with either $1$ or $\psi$ as the intermediate anyon after the second splitting process. Braiding any two of the four anyons results in the braiding matrix relating the two states:
\begin{equation}
\includegraphics[width=0.45\linewidth]{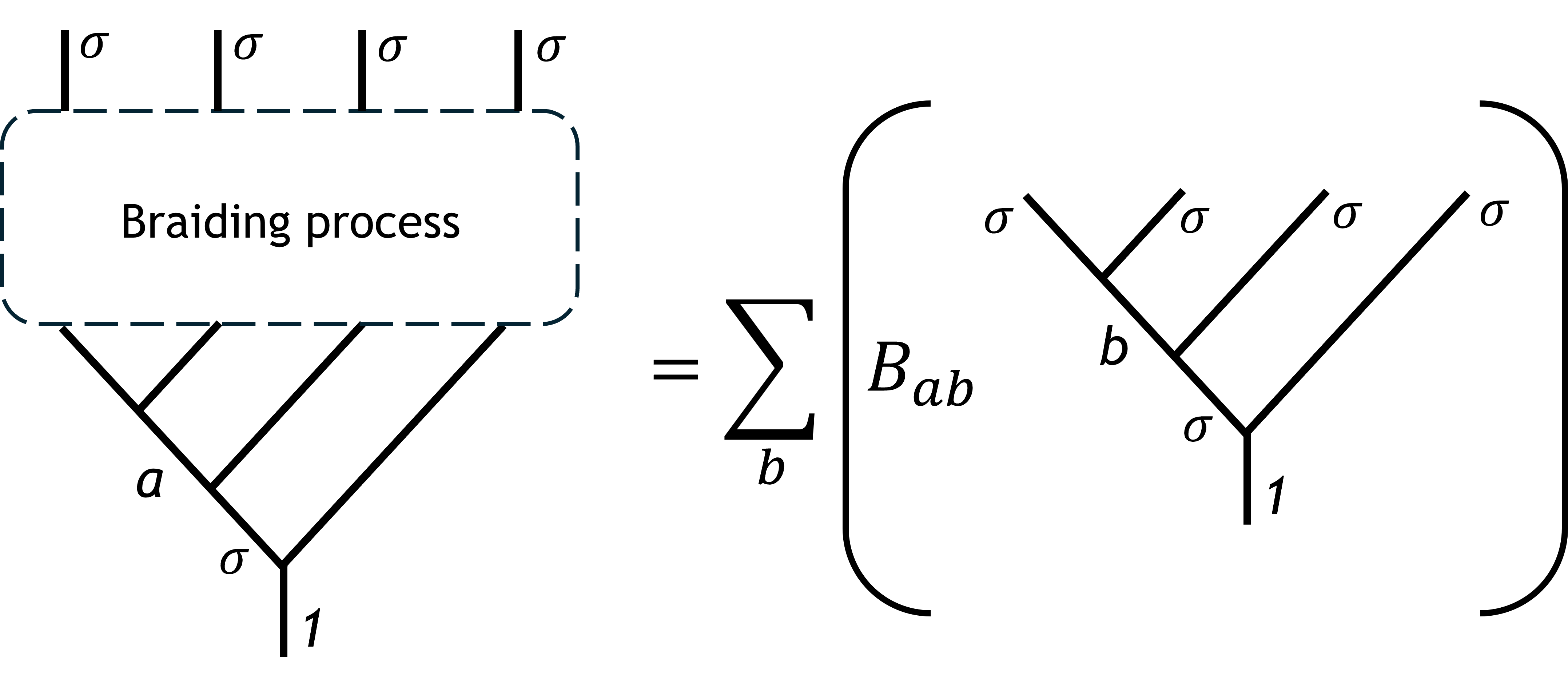}
\end{equation}
The braiding matrix can be calculated from the fusion matrices and R-matrices by resolving different subtrees to bring one diagram to another. 
\begin{comment}
To illustrate ,this we calculate the braiding matrices for one-braiding processes. There are three such processes:
\begin{equation}
\label{braiding 12}
\includegraphics[width=0.55\linewidth]{B12}
\end{equation}
\begin{equation}
\label{braiding 23}
\includegraphics[width=0.63\linewidth]{B23}
\end{equation}
\begin{equation}
\label{braiding 34}
\includegraphics[width=0.60\linewidth]{B34}
\end{equation}
We see that the processes in \Cref{braiding 12} and \Cref{braiding 34} have the same outcome:
\begin{equation}
\label{B12}
B_{ab} = R_{\sigma\sigma}^{a}\delta_{ab} \equiv e^{-i\pi/8}
\begin{pmatrix}
1 && 0 \\ 0 && i
\end{pmatrix}
\end{equation}
where as the process in \Cref{braiding 34} gives
\begin{equation}
\label{B23}
B_{ab} \equiv \frac{e^{i\pi/8}}{\sqrt{2}}
\begin{pmatrix}
1 && -i \\ -i && 1
\end{pmatrix}
\end{equation}
\end{comment}
To illustrate this process, we compute the braiding for a simple two-quasihole exchange. One can choose which two of the four "branches" in Eq. \ref{eq:four qh tree} to braid, with different choices resulting in different braiding matrices, but all related by a basis transformation. Consider first the exchange of the two left-most branches:
\begin{align}
    \includegraphics[height=80pt]{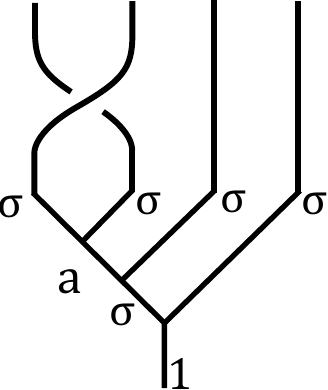}\includegraphics[height=40pt]{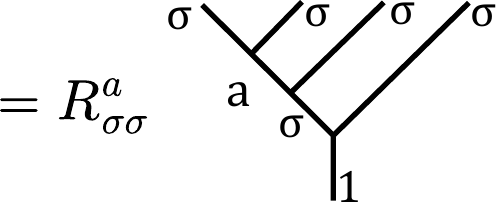}
\end{align}
In this case, the crossing can be simply resolved by a single application of \Cref{eq:R-matrix}. As a result, the braiding matrix is diagonal with the entries given by $R_{\sigma\sigma}^1$ and $R_{\sigma\sigma}^\psi$:
\begin{equation}
    \label{B12}
    B^{(12)}=e^{-i\pi/8}\begin{pmatrix}1&0\\0&i\end{pmatrix}
\end{equation}
Now, if we instead exchange the middle two branches in \Cref{eq:four qh tree}, the braiding matrix can be computed by
\begin{align}
    \includegraphics[height=80pt]{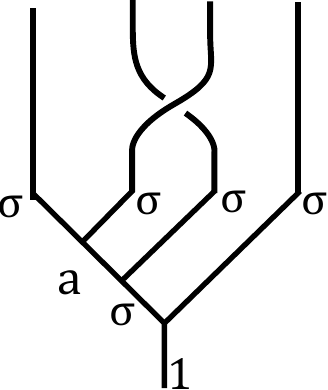}\hspace{10pt}&\includegraphics[height=80pt]{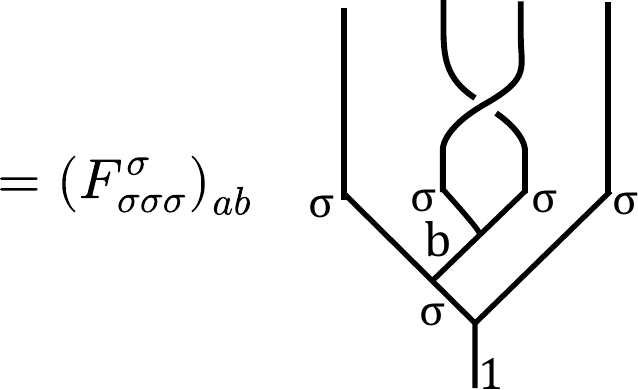}\label{derive B23 1}\\
    &\includegraphics[height=40pt]{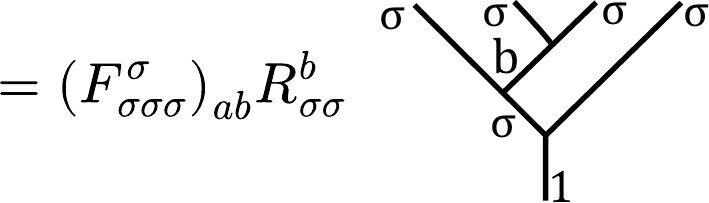}\label{derive B23 2}\\
    &\includegraphics[height=40pt]{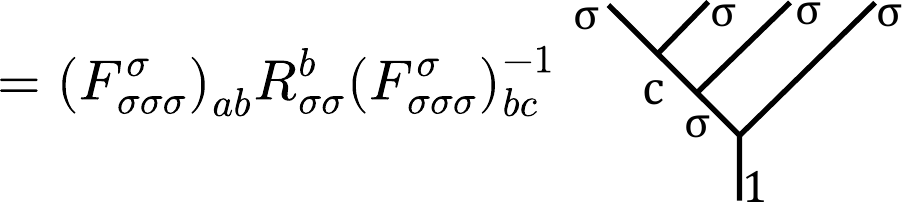}\label{derive B23 3}
\end{align}
This time, the crossing cannot simply be resolved by the R-matrix because the two anyons involved do not come from the same parent branch. However, they can be made into the same branch by the application of the fusion matrix as seen in \Cref{derive B23 1}. The crossing can be resolved afterwar,d and the creation tree can be brought back to the original tree by the inverse of the fusion matrix. The overall effect of this is the braiding matrix
\begin{equation}
    \label{B23}
    B^{(23)}=\frac{e^{i\pi/8}}{2}\begin{pmatrix}
        1+i&1-i\\
        1-i&1+i
    \end{pmatrix}
\end{equation}
which is related to \Cref{B12} simply by a basis transformation $B^{(23)}=FB^{(12)}F^{-1}$ (where $F\equiv F_{\sigma\sigma\sigma}^\sigma$). We can also compute what happens when the two right-most anyons are exchanged:
\begin{align}
    \includegraphics[height=80pt]{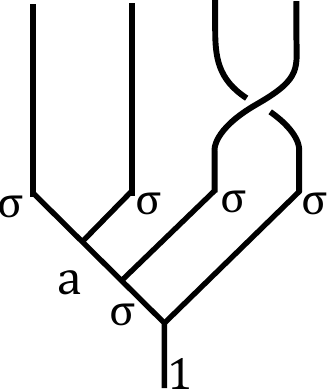}\hspace{10pt}&\includegraphics[height=80pt]{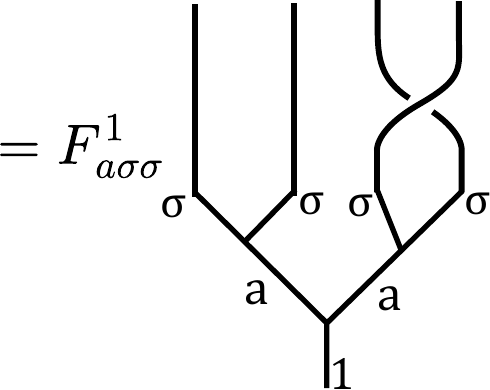}\label{derive B34 1}\\
    &\includegraphics[height=40pt]{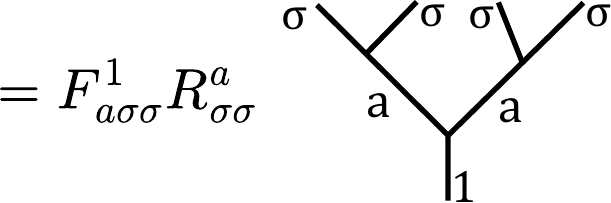}\label{derive B34 2}\\
    &\includegraphics[height=40pt]{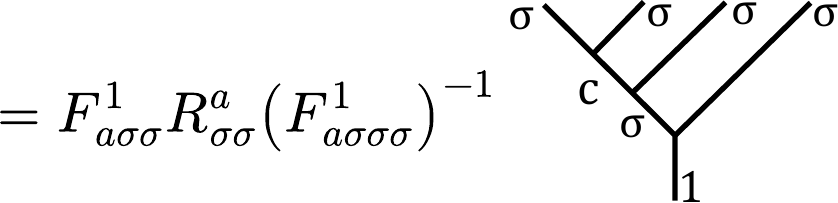}\label{derive B34 3}
\end{align}
Similar to the previous case, a fusion matrix is needed to group the two anyons involved in the crossing into the same parent branch. However, in this case, the corresponding fusion matrix is scalar: $F_{a\sigma\sigma}^1=\left(F_{a\sigma\sigma}^1\right)^{-1}=1$. Thus, the braiding matrix is identical to the first case:
\begin{equation}
    \label{B34}
    B^{(34)}=B^{(12)}
\end{equation}
Note that all of the matrices $B^{(12)}$, $B^{(23)}$, and $B^{(34)}$ describe the result of exchanging two quasiholes while leaving the other two fixed. Their difference is a result of a basis transformation, which reflects the fact that braiding matrices are \emph{not} basis-independent. The matrix $B^{(12)}$ reflects the braiding process computed numerically in \Cref{fig:braiding_result}b in the main text and agrees with \Cref{eq:braiding_matrix_result} up to a scalar phase $e^{i(\xi-\pi/2}$ and a basis transformation by $\sigma_z$ (swapping the two diagonal entries).

\subsection{Braiding Matrices of Rotational Braiding Schemes}
\begin{figure}
    \centering
    \includegraphics[width=\linewidth]{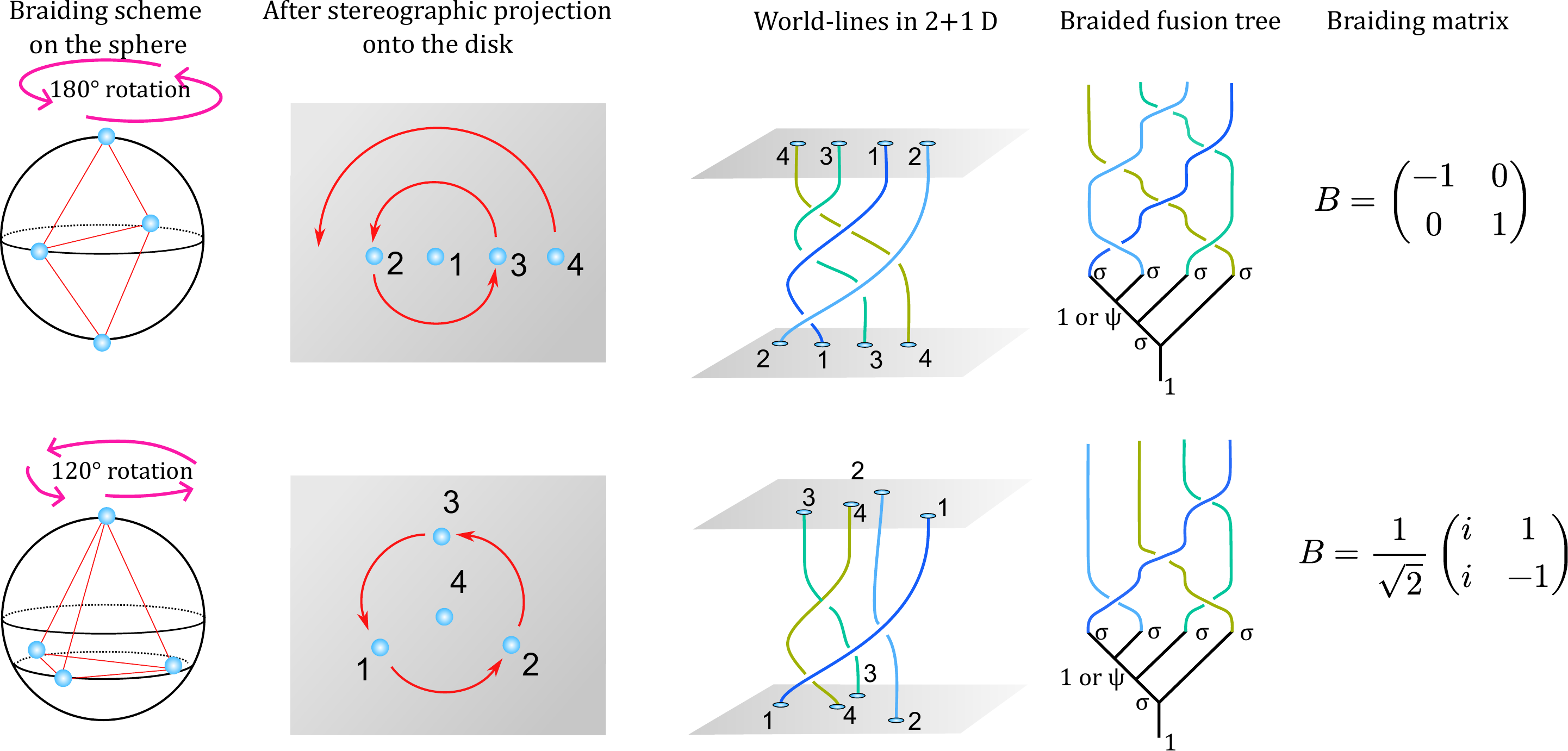}
    \caption{Each row from left to right: A schematic of the braiding process on the sphere where quasihole trajectories are generated by rotation of the entire sphere, the corresponding braiding scheme on the disk obtained from the stereographic projection (note that the north pole maps to the center of the disk and the south pole maps to infinity), the world-lines of each quasihole in this braiding scheme in 2+1 dimension, the corresponding braided fusion tree, and the braiding matrix obtained from resolving the fusion tree diagram.}
    \label{fig:rotational schemes}
\end{figure}
In this section, we detail the step-by-step procedure to compute analytically the braiding matrix of any given braiding scheme on the sphere. The first step is to map the braiding scheme on the sphere onto the disk via the stereographic projection, keeping in mind that the southpole is mapped to a point infinitely far away from the center of the disk. This means that although on the sphere, any quasihole at the south seemingly remains stationary when the sphere is rotated, in reality, it moves along a circle of infinite radius, encircling the entire system. Thus, in the first braiding scheme shown in \Cref{fig:rotational schemes}, the quasihole at the south pole moves in a semicircle around the other three when the sphere rotates by 180$^\circ$. On the disk, this process does not result in the same initial configuration (the quasihole labeled ``4" in \Cref{fig:rotational schemes} does not return to the initial position of itself or any other quasihole), but on the sphere the final configuration is the same as initial since every point along the circle of infinite radius is shrunk into the south pole. 

To see that this is indeed a valid braiding scheme, we draw the world lines of each particle, ensuring that the initial and final configurations are just a permutation of each other. A simple way to relate the picture on the second column of \Cref{fig:rotational schemes} and the picture on the third column is by imagining time as a third dimension pointing outside the page, and rotating the whole system so that our "eyes" see from the bottom of the page. Note also that from this step onward, the actual trajectory of each quasihole is not important so long as the braid structure is conserved. (Each world-line may ``wiggle" about but cannot cross over another.)

In the next step, we translate the world-line picture into a braided fusion tree. This can be done by reading off the crossings of different pairs of world-lines. There is a subtlety to be noted here about the numbering of the quasiholes. In principle, such labeling is superficial since all quasiholes are identical. Here, we label the quasiholes from 1 to 4 that corresponds to the variables $\eta_i$ in \Cref{MR 2qh state} in the main text. In our choice of form of the wave functions used as the basis for computing the matrix, the variables $\eta_1$, $\eta_2$, $\eta_3$, and $\eta_4$ correspond to the four quasiholes going from left to right on the fusion tree in \Cref{eq:four qh tree}. Thus, the crossings on the braided fusion tree must be drawn in a way that is consistent with the labeling. To facilitate this, the world line of each quasihole is colored differently in \Cref{fig:rotational schemes}. Different labeling schemes, e.g., by permutating the labels $(1,2,3,4)$ in Fig. \ref{fig:rotational schemes} correspond to choosing different bases, which would result in different braiding matrices related to one another by a basis transformation.

The braiding matrix can be computed from the tree diagram by resolving the crossing using the fusion $F$-matrix and $R$-matrix described in the previous section. Since the crossings can be resolved from bottom to top, and since we have already resolved the three single-crossing cases in the previous section, there is a simple method to read off the braiding result. We first read off the two indices $(a,b)$ (where $a$ and $b$ are integers between 1 and 4) of the two anyons involved in each crossing. Then, for each pair, we write down the braiding matrix $B^{(a,b)}$. The total braiding matrix can then be written as a product of these pair-exchange matrices, ordered from right-to-left as we move up the fusion tree. Following this procedure, the braiding matrix for the first braiding scheme (top row of \Cref{fig:rotational schemes}) is given by
\begin{equation}
    \label{scheme 1 matrix}
    B=B^{(23)}B^{(12)}B^{(34)}B^{(23)}B^{(12)}\left[B^{(34)}\right]^{-1}=\begin{pmatrix}-1&0\\0&1\end{pmatrix}
\end{equation}
Note that here the inverse of the $B^{(a,b)}$ must be used if the crossing goes ``right-over-left" instead of "left-over-right". The braiding matrix of the second rotational braiding scheme (bottom row of \Cref{fig:rotational schemes}) is
\begin{equation}
    \label{scheme 2 matrix}
    B=B^{(34)}B^{(23)}B^{(12)}\left[B^{(34)}\right]^{-1}=\frac{1}{\sqrt2}\begin{pmatrix}i&1\\i&-1\end{pmatrix}
\end{equation}
\begin{comment}
For the first braiding scheme (first row in \Cref{fig:rotational schemes}:
\begin{align}
    \includegraphics[height=120pt]{derive1-1.pdf}\hspace{10pt}&\includegraphics[height=120pt]{derive1-2.pdf}\\
    &\includegraphics[width=110pt]{derive1-3.pdf}\\
    &\includegraphics[width=166pt]{derive1-4.pdf}\\
    &\includegraphics[width=205pt]{derive1-5.pdf}
\end{align}
thus for this process we see that the result simplifies to a singular diagonal matrix $R_{aa}^1=\text{diag}(1,-1)$. For the second braiding scheme, the braiding matrix is computed as
\begin{align}
    \includegraphics[height=110pt]{derive2-1.pdf}\hspace{10pt}&\includegraphics[height=90pt]{derive2-2.pdf}\\
    &\includegraphics[height=90pt]{derive2-3.pdf}\\
    &\includegraphics[height=90pt]{derive2-4.pdf}\\
    &\includegraphics[height=48pt]{derive2-5.pdf}\\
    &\includegraphics[height=48pt]{derive2-6.pdf}\\
    &\includegraphics[height=48pt]{derive2-7.pdf}
\end{align}
The outcome is a product of matrices $R_{\sigma\sigma}^a=e^{-i\pi/8}\text{diag}(1,i)$, $F_{\sigma\sigma\sigma}^\sigma=\left(F_{\sigma\sigma\sigma}^\sigma\right)^{-1}=\frac12\begin{pmatrix}1&1\\1&-1\end{pmatrix}$, and $\left(R_{b\sigma}^\sigma\right)=\text{diag}(1,i)$. The result is as shown in \Cref{fig:rotational schemes} (\Cref{braiding matrix rotation scheme 2} in the main text.).
\end{comment}
\section{Double Stereographic projection}
\label{sec:SM_double_projection}
To map the sphere to a compact region in the complex plane while preserving the desirable properties of the standard stereographic projection, we introduce a \emph{double stereographic projection}.

In the standard (single) stereographic projection, a point on the sphere with spherical coordinates $(\theta, \phi)$ is mapped to the complex plane via $z = e^{i\phi} \tan(\theta/2)$. This mapping projects the northern hemisphere ($\theta < \pi$) onto the interior of the unit disk ($|z| < 1$), while the southern hemisphere ($\theta > \pi$) is projected to the exterior ($|z| > 1$). Critically, the south pole ($\theta = \pi$) is sent to infinity ($|z| \to \infty$). This infinite extent causes a severe problem for Monte Carlo simulations: sampling near the south pole requires integrating over an arbitrarily large region of the complex plane, even though the corresponding region on the sphere has finite area. This makes efficient and uniform sampling practically impossible in the standard projection. 

The double stereographic projection addresses this issue by combining two stereographic projections, one for each hemisphere. Concretely, the mapping is defined as

\begin{equation}
z=
\begin{cases}
  e^{i\phi} \tan(\theta/2) \;\;\text{ if }\theta\leq\pi/2 \\
  e^{i\phi} \cot(\theta/2) \;\;\text{ if }\theta>\pi/2
\end{cases}
\end{equation}
where on top of the usual mapping, an additional inversion ($w=1/z^*$) maps the exterior of the unit disk ($|z| > 1$) back into the interior of a unit disk ($|w| < 1$). As a result, the southern hemisphere is now projected onto a second unit disk, distinct from the first one that contains the northern hemisphere, hence the name double stereographic projection.

The entire sphere is thus compactly mapped onto two separate unit disks in the complex plane. This construction eliminates any points at infinity and confines the configuration space to a bounded region, making it ideal for Monte Carlo sampling.

\begin{figure}
\begin{center}
\includegraphics[width=0.5\linewidth]{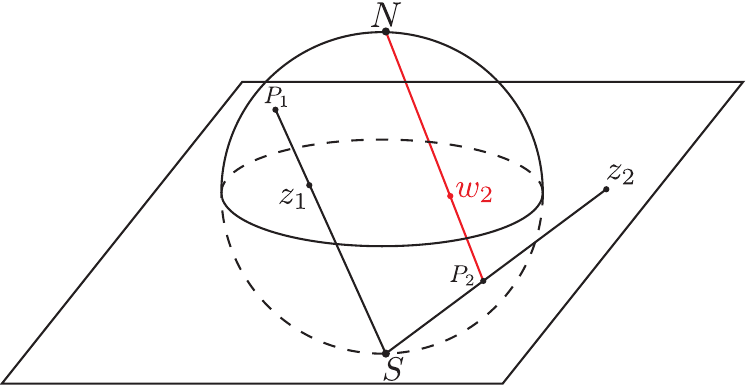}
\caption{\textbf{Double stereographic projection for HMC.} $P_1$, $P_2$ are two points on the sphere, and $z_1$, $z_2$ are points on the complex plane from the usual stereographic mapping, that are the intersections of the straight line connecting the south pole and points on the sphere with the complex plane. Since $P_2$ is located in the southern hemisphere, one has $|z_2|>1$. The double stereographic projection for $P_2$ starts from the north pole instead and maps $P_2$ to $w_2=1/z_2^*$, which is inside the unit disk.}
\label{fig:stereo_map}
\end{center}
\end{figure}

The new projection only changes the holomorphic part in the wave function and keeps the decoupled form factor term unchanged. The Laughlin wave function on a sphere under the usual stereographic projection is defined as

%%%%%%%%%%%%%%%%%%%%%%%%%%%%%%%%%%% Spinor form
% \begin{equation}
% \Psi_m(\{z_j\})=\prod_{i<j}^N(u_iv_j-v_iu_j)= \prod_{i<j}^N (z_i - z_j)^m \prod_i \left( \frac{1}{1 + |z_i|^2} \right)^S,
% \end{equation}
% where 
% \begin{equation}
% u_i=\cos\frac{\theta}{2}e^{\frac{i\phi}{2}} \text{, }\; v_i=\sin\frac{\theta}{2}e^{-\frac{i\phi}{2}}
% \end{equation}
% are the spinor coordinates.
%%%%%%%%%%%%%%%%%%%%%%%%%%%%%%%%%%% Spinor form

\begin{equation}
\Psi_m(\{z_j\})=\prod_{i<j}^N (z_i - z_j)^m \prod_i \left( \frac{1}{1 + |z_i|^2} \right)^S=\prod_{i<j}^ND_{ij}^m,
\label{eq:Laughlin_sphere}
\end{equation}
where 
\begin{equation}
D_{ij}=\frac{z_i-z_j}{\sqrt{1+|z_i|^2}\sqrt{1+|z_j|^2}}
\label{eq:P_cross}
\end{equation}
is the elementary rotational invariant cross term.

Suppose $P_j$ is in the southern hemisphere, i.e., $|z_j|>1$. Dividing $z_j$ from the numerator and denominator in Eq.~\ref{eq:P_cross} gives
\begin{equation}
D_{ij}=\frac{z_i/z_j-1}{\sqrt{1+|z_i|^2}\sqrt{1/|z_j|^2+1}}=\frac{z_iw^*_j-1}{\sqrt{1+|z_i|^2}\sqrt{1+|w_j|^2}},
\end{equation}
which has the same denominator as before, up to a local phase rotation, thus can be factored out as before, leading to the same form factor as Eq.~\ref{eq:Laughlin_sphere}. Therefore, in the new projection, one only needs to change all $z_i-z_j$ terms in the wave function to
\begin{equation}
z_i-z_j\rightarrow
\begin{cases}
  z_i-z_j \;\; \text{if both }P_i\text{ and } P_j \text{ are on the northern hemisphere}\\
  z_iw_j^*-1 \;\; \text{if }P_i\text{ is on the northern hemisphere}\text{, } P_j \text{ is on the southern hemisphere}\\
  1-w_i^*z_j \;\; \text{if }P_i\text{ is on the southern hemisphere}\text{, } P_j \text{ is on the northern hemisphere}\\
  -(w_i^*-w_j^*) \;\; \text{if both }P_i\text{ and } P_j \text{ are on the southern hemisphere}\\
\end{cases}
\end{equation}

This new projection offers several key advantages that make it particularly well-suited for efficient Hybrid Monte Carlo (HMC) simulations:

\begin{enumerate}
\item Compact sample space: Unlike the standard stereographic projection, where the south pole is mapped to infinity (requiring sampling over an unbounded region), the double stereographic projection treats both hemispheres symmetrically. Each hemisphere is mapped to the interior of a unit disk, resulting in a fully compact configuration space. This bounded domain allows straightforward and uniform sampling of particle positions on the sphere, without the need to handle infinite extents.
\item Smooth and well-behaved integrand: An alternative approach is to work directly in spherical coordinates $(\theta, \phi)$, but this introduces a Jacobian factor $\sin\theta$ in integrals over the sphere. When computing forces for the HMC dynamics, this leads to a divergent $\cot\theta$ term in the effective potential near the poles, causing numerical instabilities in the integrator and severely reducing the acceptance rate.  

In contrast, the double stereographic projection preserves a smooth form factor of the type $\frac{1}{1 + |z_i|^2}$ across both unit disks. This factor is finite and converges everywhere, varies slowly, and introduces no singularities, yielding stable forces and high acceptance rates in HMC updates.
\end{enumerate}

\section{Numerical scheme in obtaining the braiding matrix}
\label{sec:SM_braiding_mat}
We follow the numerical scheme introduced in Ref.~\cite{tserkovnyak2003monte}. The full contour is discretized into $n_s$ pieces. From this point forward, all states $\ket{\psi^{(l)}_i}$, for $i=1,2$ indexing the two model wave functions at step $l$, are considered normalized.

At each time step, the two model wave functions may not be orthogonal; a local change of basis $V$ is performed at each step. That is, $V^{(l)\dagger} B^{(l)}V^{(l)}=I$, forming a local orthonormal basis, where $B^{(l)}_{ij}=\overlap{\psi^{(l)}_i}{\psi^{(l)}_j}$, and $I$ is the identity matrix. Here, inspired by the CFT formalism in Ref.~\cite{Bonderson2011}, we choose the $V$ explicitly via $\ket{\psi_\pm}=\frac{1}{\sqrt{2\pm2|o|}}(\ket{\psi_1}\pm a\ket{\psi_2})$, where $o=\overlap{\psi^{(l)}_i}{\psi^{(l)}_j}$, and $a=\frac{o^*}{|o|}$. This choice produces the nearly diagonal matrix without doing any additional change of basis, other than an overall phase to the whole Berry matrix.

The inter-step overlap $W^{(l)}=\overlap{\psi^{(l)}_i}{\psi^{(l+1)}_j}$ measures how the states evolve when the quasiholes' positions are moved. Defining $A^{(l)}_{ij}=V^{(l)\dagger}W^{(l)}V^{(l+1)}+h.c.$, the infinitesimal braiding matrix accumulating at each step is computed as  $U^{(l+1)}=U^{(l)}(I+A^{(l)}/2)(I-A^{(l)}/2)^{-1}$, which can be shown to be exactly unitary at each step.

In general, the final wave functions $\ket{\psi^{(n_s)}_i}$ may not be the same as $\ket{\psi^{(0)}_i}$, although they share the same set of quasiholes coordinates and span the same Hilbert space. To keep the computation gauge independent, i.e., independent of any arbitrary change of basis during the process, a final projection is done $U^{(n_s)}\rightarrow U^{(n_s)}O^T$, where $O=V^{(0)}\Omega V^{(n_s)}$ connect the two Hilbert spaces and $\Omega_{ij}=\overlap{\psi^{(0)}_i}{\psi^{(n_s)}_j}$, is the overlap between the initial and final wave functions.

\end{document}